\documentclass[apj]{emulateapj}
\usepackage{multirow}
\usepackage{natbib}
\usepackage{graphicx}
\usepackage{epstopdf}

\begin{document}

\title{The Magnetic Field of Cloud~3 in L204}
\author{Lauren R. Cashman \& D. P. Clemens}
\affil{Institute for Astrophysical Research}
\affil{Boston University, 725 Commonwealth Ave, Boston, MA 02215}
\email{lcashman@bu.edu, clemens@bu.edu}

\slugcomment{Accepted for publication in The Astrophysical Journal}

\shorttitle{L204 Magnetic Field}
\shortauthors{Cashman \& Clemens}

\begin{abstract}

The L204 dark cloud complex is a nearby filamentary structure in Ophiuchus North that has no signs of active star formation. Past studies show that L204 is interacting with the nearby runaway O star, $\zeta$~Oph, and hosts a magnetic field that is coherent across parsec-length scales. Near-infrared $H$-band (1.6$\mu$m) linear polarization measurements were obtained for 3,896 background stars across a $1\degr \times 1.5\degr$ region centered on the dense Cloud~3 in L204, using the Mimir near-infrared instrument on the 1.8m Perkins Telescope. Analysis of these observations reveals both large-scale properties and small-scale changes in the magnetic field direction in Cloud~3. In the northern and western $\zeta$~Oph facing regions of the cloud, the magnetic field appears to be pushed up against the face of the cloud. This may indicate that the UV flux from $\zeta$~Oph has compressed the magnetic field on the western edge of L204. The plane-of-sky magnetic field strength is estimated to be $\sim 11 - 26$~$\mu$G using the Chandrasekhar-Fermi method. The polarimetry data also reveal that the polarization efficiency (PE $\equiv P_{\rm H}/A_{\rm V}$) steadily decreases with distance from $\zeta$~Oph ($-0.09 \pm 0.03~\% \, {\rm mag}^{-1} \, {\rm pc}^{-1}$). Additionally, power-law fits of PE versus $A_{\rm V}$ for localized samples of probe stars show steeper negative indices with distance from $\zeta$~Oph. Both findings highlight the importance of external illumination, here from $\zeta$~Oph, in aligning dust grains to embedded magnetic fields.

\end{abstract}


\section{Introduction}
How are molecular clouds influenced by magnetic fields? Do magnetic fields affect cloud collapse and morphology? Before stars form, molecular clouds and the dense cores within them must form. These processes, however, are not yet well understood. \citet{Crutcher12} comprehensively reviewed magnetic fields in molecular clouds, concluding that the relative contributions of turbulence, magnetic fields, and gravitational potential to cloud formation and collapse remain unknown. Star-formation theories fall mainly into two magnetic classes: strong-field and weak-field. In strong-field models, cloud collapse occurs primarily along the field lines (e.g., \citealt{Mouschovias99}). This leads to cloud long-axes predominantly perpendicular to their magnetic fields. In weak-field models, turbulent flows drive formation of clouds (e.g., \citealt{MacLow04}) and as clouds collapse, gas drags magnetic fields along, creating pinched or ``hourglass'' morphologies.

Magnetic fields may be observed, and different morphologies tested, through near-infrared (NIR) polarimetry of background starlight. The partial linear polarization of optical starlight was discovered by \citet{Hall49} and  \citet{Hiltner49a,Hiltner49b}. This polarization is attributed to dichroic extinction by aligned, asymmetric dust grains. Spinning dust grains become aligned with their long axes preferentially perpendicular to the magnetic field, resulting in partial polarization of background starlight oriented parallel to the magnetic field. The exact alignment mechanism, however, is still unknown. The current leading theory to explain how dust grains become aligned in the presence of a magnetic field is through radiative aligned torques (RATs; \citealt{Lazarian07}). RAT theory relies on an anisotropic radiation field to spin-up dust grains possessing net helicity and to align the grains with the magnetic field. Therefore, as the opacity to the illuminator increases, there is less radiation available to spin-up dust grains and grain alignment efficiency should decrease.

Differences in grain alignment efficiency can be measured through differences in polarization efficiency (PE), the ratio of polarization percentage $P_{\lambda}$ to optical depth or extinction ($A_{\rm V}$). Previous studies have shown a decrease in PE with increasing optical depth \citep{Gerakines95,Whittet08}, which is consistent with predictions made by RAT theory.

Other tests of RAT theory examine the peak wavelength of polarization ($\lambda_{MAX}$), which corresponds to the size of the dust grains being aligned. \citet{Andersson07} searched for correlations between $\lambda_{MAX}$ and the illuminating radiation field strength. They found that $\lambda_{MAX}$ increases linearly with $A_{\rm V}$, concluding that smaller dust grains become unaligned at larger optical depths. Another test, by \citet{Andersson10}, found that $\lambda_{MAX}$ becomes shorter in the presence of a young B9/A0 star, indicating that small dust grain alignment becomes more efficient near such a blue illuminator. \citet{Matsumura11} showed evidence that grain alignment efficiency is also enhanced at higher dust temperatures.

Another key point from \citet{Andersson07} is that $A_{\rm V}$ represents the line-of-sight (L.O.S.) extinction between some background star and the observer and not necessarily the effective extinction to the illuminator seen by the dust grains. They used $I(60 \mu {\rm m})/I(100 \mu {\rm m})$ as a measure of dust grain temperature, which is correlated with the radiation intensity seen by the grains. Thus, \citet{Andersson07} were able to select lines-of-sight where the observed extinction corresponded to the actual optical depth seen by the grains.

Here, use of the Mimir \citep{Clemens07} wide-field instrument for new near-infrared (NIR) polarization measurements of background starlight towards Cloud~3 in L204 is reported. These new observations had finer spatial sampling than previous L204 studies and so reveal small scale changes in the plane-of-sky magnetic field. Additionally, differences in polarization (grain alignment) efficiency throughout the cloud and with distance from the nearby O star, $\zeta$~Oph, were examined. Section 2 introduces the L204 laboratory and Cloud~3 region. The observations and data reduction are described in Section 3. Section 4 explains the analysis of the data. In Section 5, the results of the analyses and their implications are discussed. This work is summarized in Section 6.

\begin{figure*}
	\centering
	\includegraphics[width=0.8\textwidth, angle=0, trim = 0.5cm 3.9cm 0.9cm 0.9cm, clip=true]{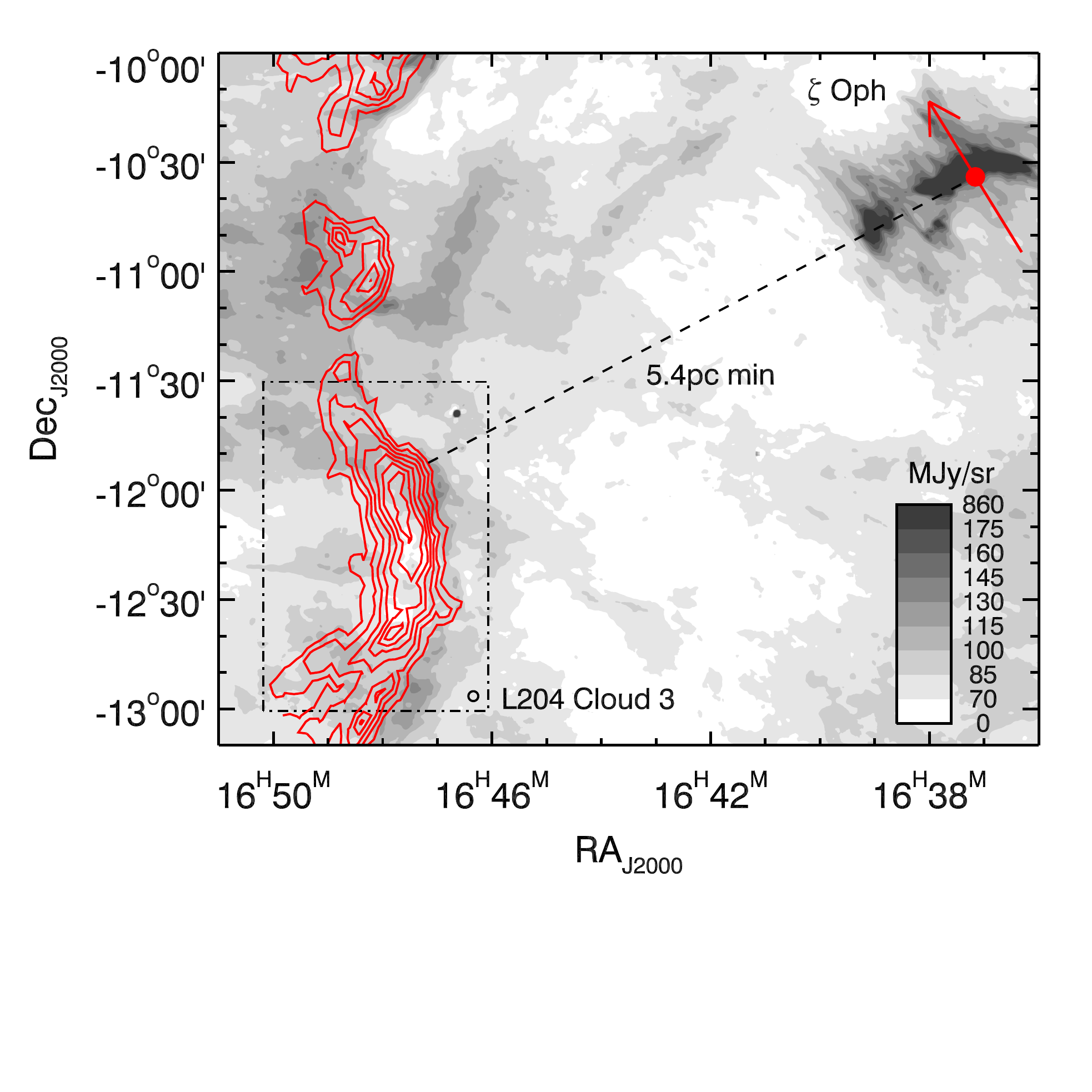}
	\caption{\label{lab} Smoothed, grayscale ${\it WISE}$ \citep{Wright10} 12$\mu$m dust map computed by \citet{Meisner14} of the L204 region. Dust emission contours are stepped linearly from 0 to 175 MJy/sr. The maximum emission in the region is 860 MJy/sr. The red contours represent $^{13}$CO integrated intensities from \citet{Tachihara00a}, ranging linearly from 1 to 7 K km s$^{-1}$. The filled red circle indicates the location of $\zeta$~Oph, with the red arrow representing its proper motion over $10^5$ years. The black dot-dashed box indicates the $1\degr \times 1.5\degr$ region for which NIR linear polarimetry measurements were obtained. (A color version of this figure is available in the online version.)}
\end{figure*}

\section{The L204 Laboratory}

The nearby dark cloud complex L204 \citep{Lynds62} provides a good laboratory to investigate both magnetic field morphology and dust grain alignment efficiency. It is a $4\degr$ long, filamentary molecular cloud in Ophiucus ($\ell = 6.6\degr$, $b = 20.6\degr$) that shows no evidence of star formation \citep{Hatchell12}. L204 appears in absorption in front of a cloud of $\rm H\alpha$ emission associated with $\zeta$~Oph \citep{Sivan74}, a runaway O9.5 V star \citep{Lesh68}. This provides an upper limit on the distance to L204 of $112.2 \pm 2.5$~pc based on the trigonometric parallax of $\zeta$~Oph \citep{Leeuwen07}. L204 and $\zeta$~Oph are separated by $\sim 2.8\degr$ on the sky, which corresponds to a distance of 5.5~pc if L204 is at the distance of $\zeta$~Oph.

\citet{McCutcheon86} observed optical polarization of background starlight for 60 stars in the L204 region. They found that the magnetic field of L204 is predominantly perpendicular to the long axis of the cloud across several degrees on the sky. The measured polarizations were also generally aligned with the mean polarization position angle in the surrounding ISM, leading \citet{McCutcheon86} to conclude that L204 collapsed along an existing magnetic field. Their CO observations revealed a steep density gradient on the western, $\zeta$~Oph facing, edge of the cloud, implying interaction with that star. They also found the velocity structure and morphological structure of the cloud are correlated. The radial velocities of the less massive sections of the cloud are consistently offset from those of the more massive sections. This correlation implies that some external force is acting on L204, accelerating the less massive sections of the cloud. Based on the assumption that the cloud is in virial equilibrium, \citet{McCutcheon86} estimated a magnetic field strength of $\sim$50~$\mu$G.

The L.O.S. magnetic field strength ($B_\parallel$) in L204 was measured at several positions by \citet{Heiles88}, using Zeeman splitting of the 21 cm line observed with the Hat Creek 85 foot telescope. \citet{Heiles88} found a correlation of $B_\parallel$ with $V_{LSR}$, which he concluded was indicative of gas flowing along the magnetic field. In that case, the magnetic field would be approximately constant throughout the cloud and differences in $B_\parallel$ would indicate variations in the direction of the magnetic field. While the maximum $B_\parallel$ measured was 9.1~$\mu$G, Heiles favored a total field strength of $\sim$12~$\mu$G. No other Zeeman observation of L204 exist.

L204 was also observed by \citet{Tachihara00a,Tachihara00b} in mm lines of $^{12}$CO, $^{13}$CO, and C$^{18}$O using NANTEN at angular resolutions as fine as 2 arcmin. They separated L204 into 6 clouds, the largest of which they called `Cloud~3.' It has a semi-major axis of 6.5~pc and a semi-minor axis of 1.5~pc, at their assumed distance of 140~pc. They estimate $M = 430M_{\odot}$ for Cloud~3, which corresponds to $M \simeq 275M_{\odot}$ at 112~pc. \citet{Tachihara00a} saw the same gas structure as \citet{McCutcheon86} and also found that the denser cores (traced by $^{13}$CO and C$^{18}$O) are on the side of the cloud closest to $\zeta$~Oph. Additionally, \citet{Tachihara00a} examined the possible effects on L204 from the stellar wind and UV flux of $\zeta$~Oph. Based on kinetic energy calculations, they concluded that momentum injection from UV photo-dissociation has the largest effect on the velocity structure of the cloud and that the stellar wind from $\zeta$~Oph has only a negligible effect.

The L204 region is shown in Figure~\ref{lab}. The red contours represent $^{13}$CO integrated intensities of the L204 cloud complex from \citet{Tachihara00a}. A solid red circle indicates the location of $\zeta$~Oph, with the red line representing its proper motion over $10^5$ years. The background grayscale image is the ${\it Wide}$-${\it field}$ ${\it Infrared}$ ${\it Survey}$ ${\it Explorer}$ (${\it WISE}$; \citealt{Wright10}) 12$\mu$m dust map computed by \citet{Meisner14}. The map reveals both the almost $1\degr$ wide bow shock in front of $\zeta$~Oph as well as a $3\degr$ radius cavity centered on $\zeta$~Oph and with its edge at L204. The black dot-dashed box indicates the $1\degr \times 1.5\degr$ region observed for this work.

\section{Observations and Data Reduction}

Linear polarimetry measurements were obtained in the NIR $H$-band (1.6$\mu$m) using the Mimir instrument \citep{Clemens07} on the 1.8m Perkins telescope outside of Flagstaff, AZ. Mimir used a 1024$^{2}$ Aladdin III InSb detector array, operated at 33.5K. The instrument had a field of view (FOV) of $10\arcmin \times 10\arcmin$ and pixel scale of $0.58\arcsec$. Polarimetric analysis was performed with a cold, stepping, $H$-band, compound half-wave plate (HWP) plus a fixed, cold wire-grid analyzer.

Sky observations were planned for a grid of 54 abutting fields (a $6 \times 9$ field grid) covering a $1\degr \times 1.5\degr$ region of L204 centered on Cloud~3. They were conducted over several nights between 2010 May and 2012 May. To obtain polarimetry measurements, exposures of 5s duration were obtained for 16 unique HWP position angles at six sky-dither positions, yielding a total of 96 images per field. The total on-sky integration time of the entire map (8 minutes per FOV) was 7.2 hours. Polarization flat-fields for each HWP position were taken using a lights-on/lights-off method against a uniformly illuminated flat-field screen inside the closed telescope dome.

Polarimetric standard stars from the list of \citet{Whittet92} and globular cluster stars were observed on the same nights to determine the instrumental polarization, polarimetric efficiency, and HWP position angle offset.

Polarization data were reduced and analyzed using custom IDL software, as described in \citet{Clemens12b}, yielding Stokes $U$ and $Q$, polarization $P$, equatorial position angle P.A., and their uncertainties, for 7,150 stars contained within the surveyed region. The biased degree of polarization ($P'$) and equatorial position angle (P.A.) were calculated from the Stokes parameters:
\begin{equation}
	\label{p_equ}
	P' = \sqrt{U^2 + Q^2}
\end{equation}
\begin{equation}
	\label{pa_equ}
	\mathrm{P.A.} = \frac{1}{2} \, \mathrm{arctan}\left(\frac{U}{Q}\right)
\end{equation}
where P.A. is in radians. The degree of polarization ($P'$) was then Ricean-corrected, to account for positive bias, following the procedure of \citet{Wardle74}:
\begin{equation}
	\label{Ricean_P}
	P = \sqrt{P'^2 - \sigma_{\rm P}^2}
\end{equation}
where $\sigma_{\rm P}$ is the uncertainty in the polarization percentage. From these quantities, the uncertainty in the polarization position angle $\sigma_{\rm P.A.}$, in radians, was calculated:
\begin{equation}
	\label{pa_uncert}
	\sigma_{\rm P.A.} = \frac{1}{2} \left(\frac{\sigma_{\rm P}}{P}\right); P \neq 0.
\end{equation}

\begin{figure}
	\centering
	\includegraphics[angle=0, trim = 1.5cm 0.9cm 2.7cm 0.7cm, clip=true, scale=0.49]{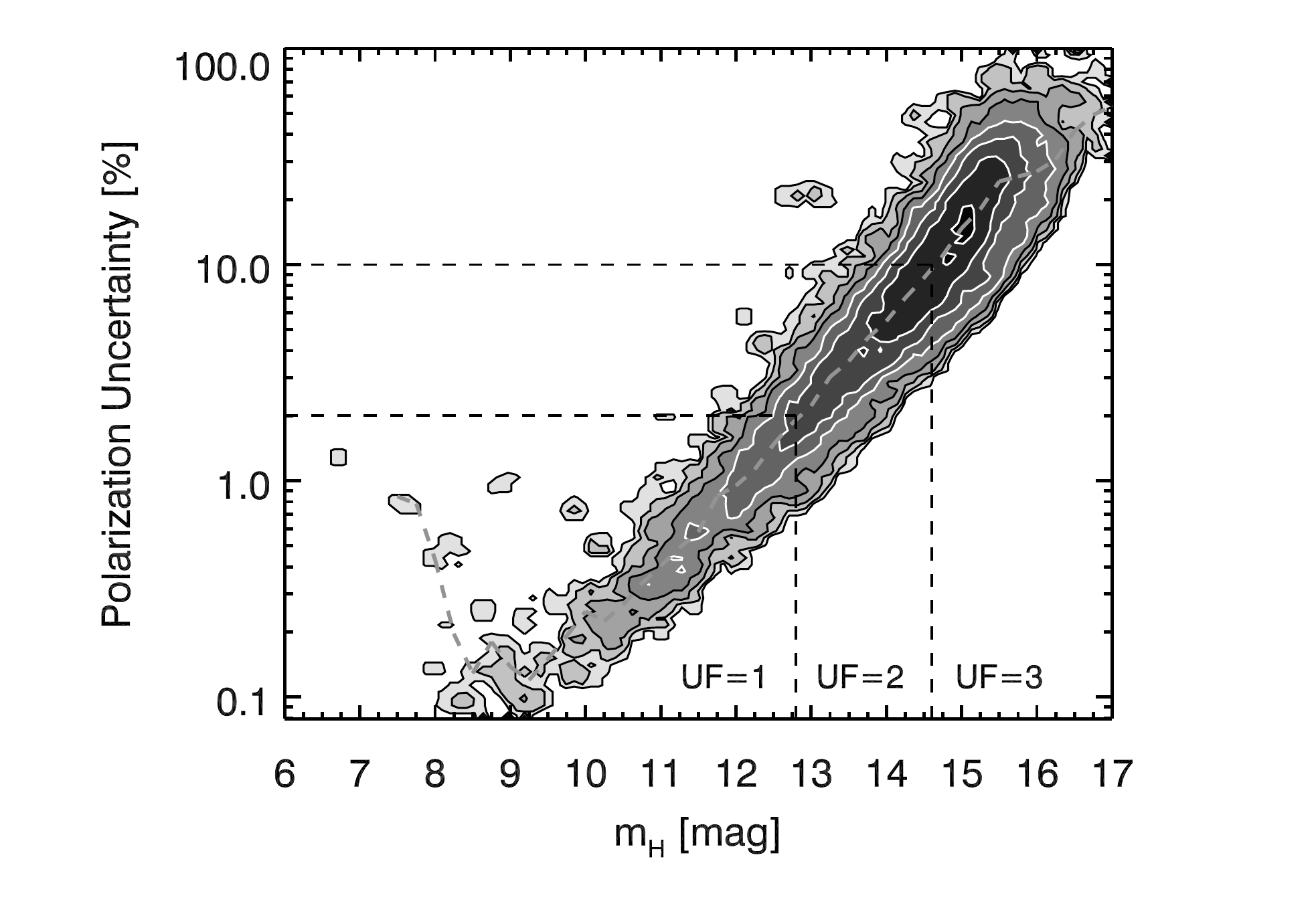}
	\caption{\label{UF} Uncertainty of polarization percentage $\sigma_{\rm P}$ for L204 region stars vs. their Mimir observed $H$-band magnitudes $m_{H}$. Shown are filled contours from 0.70\% to 90\% of the maximum of stellar density, scaled by a factor of two between each step. The three Usage Flag (UF) regions are identiﬁed, as described
	in the text. A dashed gray curve traces the run of median values of $\sigma_{\rm P}$, computed for each 0.25 mag wide band of $m_{H}$.}
\end{figure}

The stars were separated into three samples, indicated by Usage Flags \citep[UFs:][]{Clemens12a} for each star, as shown in Figure~\ref{UF}. Stars with UF = 1 have the highest polarimetric quality, characterized by $\sigma_{\rm P} \leq 2\%$ and $m_{H} \leq 12.8$ mag. The lower quality UF = 2 sample is comprised of those stars outside of the UF = 1 region, but with $\sigma_{\rm P} \leq 10\%$ and $m_{H} \leq 14.6$ mag. Stars with UF = 3 are the lowest quality sample and lay outside of both the UF = 1 and 2 regions. The boundary magnitudes were determined by the $H$-band magnitudes at which the median $\sigma_{\rm P}$ were equal to 2\% and 10\%. As shown by \citet{Clemens12a}, although the UF = 2 and 3 stars have lower polarimetric SNR, their Stokes parameters can be averaged spatially (with variance weighting) to provide useful information.

Table~\ref{stars} lists the Mimir observed polarization properties and 2MASS \citep{Skrutskie06} photometric magnitudes (where available) for 7,150 stars with polarimetric measurements, ordered by RA. Column 1 lists the star number and columns 2 and 3 list the RA and decl. The Mimir measured quantities Stokes $U$ and $Q$, polarization $P$, equatorial position angle P.A., and UF are given in columns 4 through 8, respectively. For the 6,476 that were matched to 2MASS, their $J$, $H$, and $K$-band magnitudes are listed in columns 9-11. For the remaining 674 stars with no 2MASS match, these columns are set to  $99.99 \pm 0.00$ mag. Columns 12 and 13 present the estimated visual extinction and Background Flag (BF), described in Section 4.1.

\begin{figure}
	\centering
	\includegraphics[width=0.45\textwidth,angle=0, trim = 1.0cm 0cm 3.3cm 0.9cm, clip=true]{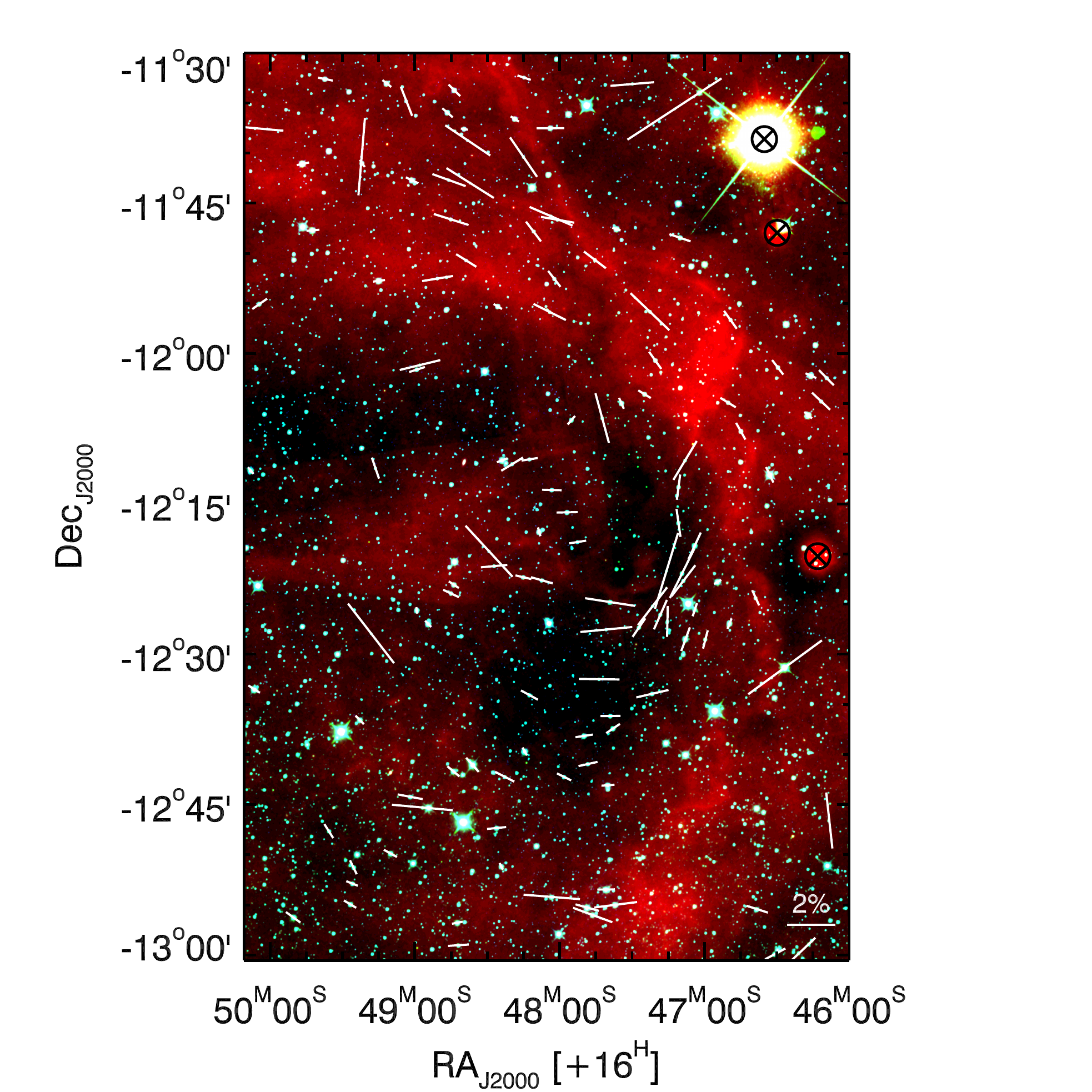}
	\caption{\label{WISE} ${\it WISE}$ three-color image of the Mimir surveyed region, constructed from 3.4$\mu$m (blue), 4.6$\mu$m (green), and 12$\mu$m (red) data. The 3.4$\mu$m and 4.6$\mu$m bands trace the stellar population while 12$\mu$m emission traces warm dust in the region. Linear polarizations of UF = 1 stars also having $P/\sigma_{\rm P} \geq 2.5$ are shown as white vectors. Vector lengths represent percentage polarization $P$, with a reference vector shown in the lower right. Vector angles represent the polarization position angle, P.A., measured East from North in this equatorial representation in J2000 coordinates. Three black circles with `X's represent saturated pixels and persistences in the ${\it WISE}$ images. The upper right bright star (with circled `X') is V446~Oph, discussed in Section 5.2.1. (A color version of this figure is available in the online version.)}
\end{figure}

Figure~\ref{WISE} shows a ${\it WISE}$ three-color image of the Cloud~3 region observed with Mimir. The ${\it WISE}$ image shows that the opaque core of Cloud~3 is rimmed by a filament of warm dust on the western, $\zeta$~Oph facing, edge. White vectors represent polarizations of the 188 stars having both UF = 1 and $P/\sigma_{\rm P} \geq$ 2.5 ($\sigma_{\rm P.A.} \leq 11.5\degr$). The polarization percentages $P$ are encoded as vector lengths and the P.A.s as vector orientations. A 2\% reference vector is shown in the bottom right corner. The oversaturated star at RA = 16:46:39.11 decl. = $-$11:38:53.1 is V446~Oph and will be discussed in Section 5.2.

\section{Analysis}

\subsection{NICER Extinctions and Map}
Due to the nature of background starlight polarimetry, only stars that are behind L204, and whose polarizations can be attributed to the cloud, are useful for magnetic field analyses. To establish the background status of the stars in the field, visual extinctions were estimated to each of the 6,476 stars having a 2MASS match using the Near-Infrared Color Excess Revisited (NICER) technique \citep{Lombardi01}.

The NICER method\footnote{Using the C. Beaumont IDL implementation.} uses NIR colors to estimate extinctions for a region based on comparison to the intrinsic colors of a nearby unextincted stellar population. An existing extinction map of the Ophiuchus cloud from \citet{Lombardi08} and ${\it WISE}$ 12$\mu$m and 22$\mu$m images were used to identify a control field of negligible extinction and dust emission. The control field is centered at RA = 16:55:45.3, decl. = $-$11:34:58 and has a radius of 35~arcmin. All 2MASS stars in the control field having $\sigma_{J}$, $\sigma_{H}$, $\sigma_{K} \leq 0.25$ mag were used to calculate the intrinsic colors and color scatter of these presumed unextincted stars. These colors were compared to those in the Cloud~3 region to estimate their visual extinctions (listed in Table~\ref{stars}).

\begin{figure}
	\centering
	\includegraphics[width=0.45\textwidth, angle=0, trim = 2.5cm 0.8cm 1.6cm 1.8cm, clip=true]{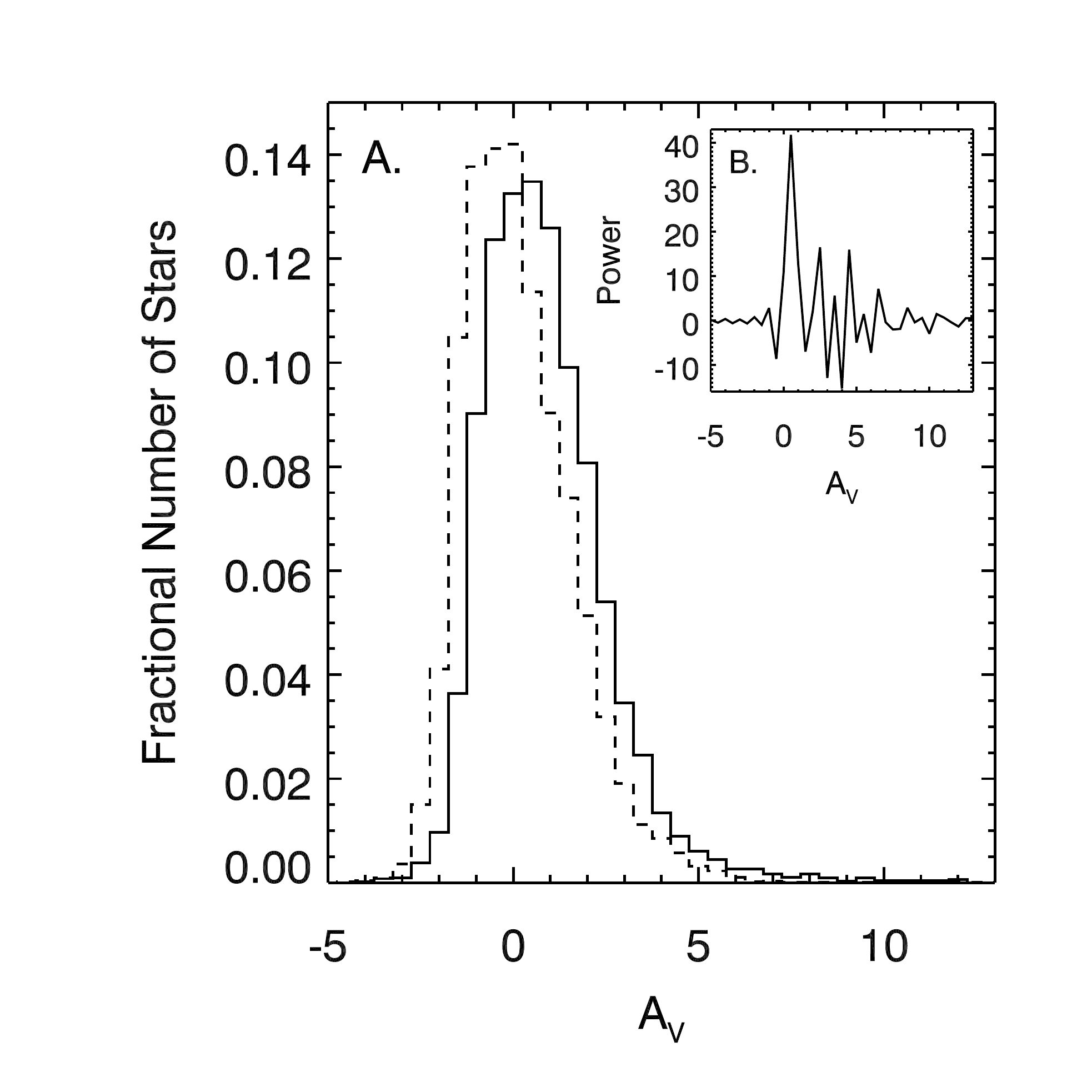}
	\caption{\label{histogram} (A.) Histogram of NICER extinctions for the Cloud~3 stars (solid) and the control field stars (dashed). Both histograms have been normalized so that their total areas are unity. The stars along the line of sight towards Cloud~3 reveal a single layer of extinction, which peaks at $A_{\rm V} \sim 0.5$ mag. (B.) Inset plot in the upper right shows the extinction power spectrum of L204, corrected for the NICER extinctions of the control region to remove the effects of the intrinsic scatter of stellar colors. It also peaks at 0.5 mag and reveals a lack of any comparable secondary extinction layer.}
\end{figure}

A histogram of these estimated Cloud~3 extinctions is shown in Figure~\ref{histogram}. The solid histogram shows the distribution of NICER extinctions for stars in the Cloud~3 region. The dashed histogram shows the distribution of NICER extinctions applied to the stars in the control field. As a result of the large spread of intrinsic colors, the NICER extinctions for both the control field and Cloud~3 region have wide distributions and extend to negative values. These negative values are allowed by the NICER analysis and indicate negligible extinction along the line of sight. As expected, the control sample has a distribution peaked close to $A_{\rm V}$ = 0 mag. The extincted distribution peaks at $A_{\rm V} \sim 0.5$ mag and may have a weaker higher extinction tail.

To constrain which stars are most likely to be behind Cloud~3, the signature of the unextincted stellar population in the control field needed to be removed from that of the Cloud~3 background stellar population. This was done by direct deconvolution of the distributions shown in Figure~\ref{histogram}. Fast Fourier transforms were computed of the histograms of both populations and then ratioed. The inverse fast Fourier transform of the ratio revealed the extinction power spectrum as a function of $A_{\rm V}$, shown as the inset in Figure~\ref{histogram}. The power spectrum peaks at $A_{\rm V}=0.5$ mag, with an amplitude $5.7\sigma$ above the neighboring noise, and no other peak rises above $2.5\sigma$. Thus, Cloud~3 contributes 0.5 mag of $A_{\rm V}$ on average. Any star showing NICER $A_{\rm V} \geq 0.5$ mag is deemed to be behind the cloud and has its Table~\ref{stars} BF value set to 2 to indicate its background status.

Of the 6,476 stars measured polarimetrically and matched to 2MASS, 3,896 have $A_{\rm V} \geq 0.5$ mag and so were judged to be background (BF = 2) while the remaining 2,580 are foreground (BF = 1). No visual extinction estimate could be made for the 674 stars with no 2MASS match, and are therefore of uncertain location (BF = 0).

Due to the proximity of L204 and its large angular distance from the Galactic mid-plane ($b=20.6\degr$), all of the extinction along the line of sight is likely due to L204. \citet{McCutcheon86} estimated that only 0.3 mag of extinction is contributed by foreground material and that the total $A_{\rm V}$ along a line of sight to the edge of the Galaxy at $b = 20\degr$, if there are no molecular clouds, is $\sim$0.6 mag. Our conclusion of an average 0.5 mag of $A_{\rm V}$ attributable to Cloud 3 is consistent with their estimates.

\begin{figure}
	\centering
	\includegraphics[width=0.45\textwidth, angle=0, trim = 1.0cm 0.5cm 4.1cm 1.1cm, clip=true]{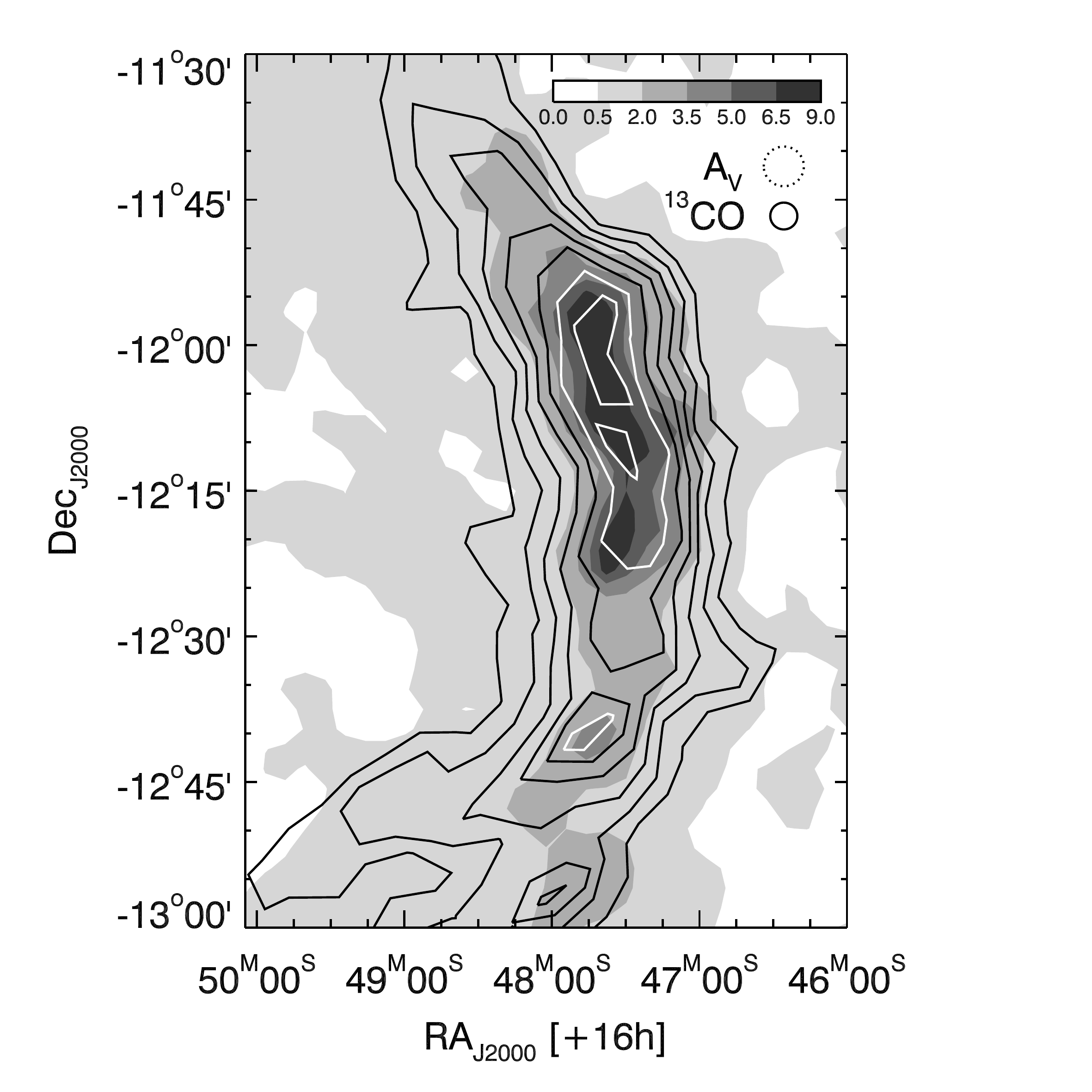}
	\caption{\label{extinction} NICER extinction map across the Mimir surveyed region. Extinctions are displayed as gray filled regions with levels of 0.5 mag through 6.5 mag, stepped linearly, then 9.0 mag as shown in the scale bar in the upper right. Black and white contour lines represent $^{13}$CO integrated intensities from Figure~\ref{lab} \citep{Tachihara00a}. The dotted circle represents the FWHM of the Gaussian smoothing kernel used to create the extinction map. The solid circle represents the Half-Power Beam Width of the $^{13}$CO observations.}
\end{figure}

To create an extinction map, stars were collected into a rectangular grid of circular, overlapping bins with radii of $8.5\arcmin$ and separations of $2\arcmin$. Weighted means were calculated within each bin from the individual stellar (NICER) extinctions, their uncertainty-based variances, and a spatial Gaussian weighting kernel having FWHM of $4\arcmin$. The FWHM and bins were chosen to meet the Nyquist sampling limit and have $\langle A_{\rm V} \rangle$ SNR $\geq 3$ over $50\%$ of the grid points. The resulting extinction map is shown in Figure~\ref{extinction} as a grayscale image, which ranges from $A_{\rm V}$ = 0 to 9 mags. Overlaid as black and white contours are the same $^{13}$CO integrated intensities from Figure~\ref{lab} \citep{Tachihara00a} to show the location of Cloud~3. As expected, regions of higher visual extinction correspond to high $^{13}$CO column densities and the dust extinction structure closely matches that of the gas. Additionally, the regions of lower extinction ($A_{\rm V}$ = 0.5 - 2 mags) closely match the extended 12$\mu$m warm dust distribution seen in Figure~\ref{WISE}.

To test strong-field models, it is necessary to identify the long axis of the cloud. The NICER extinction map in Figure~\ref{extinction} reveals the curve of Cloud~3, as seen by \citet{McCutcheon86} and \citet{Tachihara00a}, which must be taken into account when determining the location of the long axis. The extinction map was used to locate the locus of peak local extinction of the cloud (i.e., the cloud `ridge'). The RA locations of the peak extinction found in decl. strips in the map were fit with a least-squares polynomial. An F-test supported, third-order polynomial fit was found to best describe the spatial variation of the ridge. This ridge locus is plotted in later figures and is featured in the analyses described below.

\subsection{Magnetic Field Morphology}
As seen in Figure~\ref{WISE}, the highest-significance polarization stars are sparsely distributed and do not probe the most opaque parts of Cloud~3. To increase the polarization SNR across the whole cloud and to reach into the opaque cores, variance-weighted mean values of Stokes $U$ and $Q$ were calculated using all 3,896 BF = 2 (i.e., background) stars having measured polarizations (i.e., including stars of all UF values). Stokes $U$ and $Q$ maps were generated using the same gridding technique employed for the NICER extinction map. The FWHM of the spatial Gaussian kernel was $6.6\arcmin$ and the bins had radii of $11\arcmin$ ($4\sigma$ of kernel) and center separations of $3.3\arcmin$ (HWHM of the kernel). The kernel and bins were chosen to meet the Nyquist sampling limit and retain $P/\sigma_{\rm P} \geq$ 3 across 50\% of the map area. The Stokes parameter maps were used to determine maps of $P$ and P.A., as well as their uncertainties.

\begin{figure}
	\centering
	\includegraphics[width=0.45\textwidth, angle=0, trim = 0.6cm 0cm 3.6cm 0.8cm, clip=true]{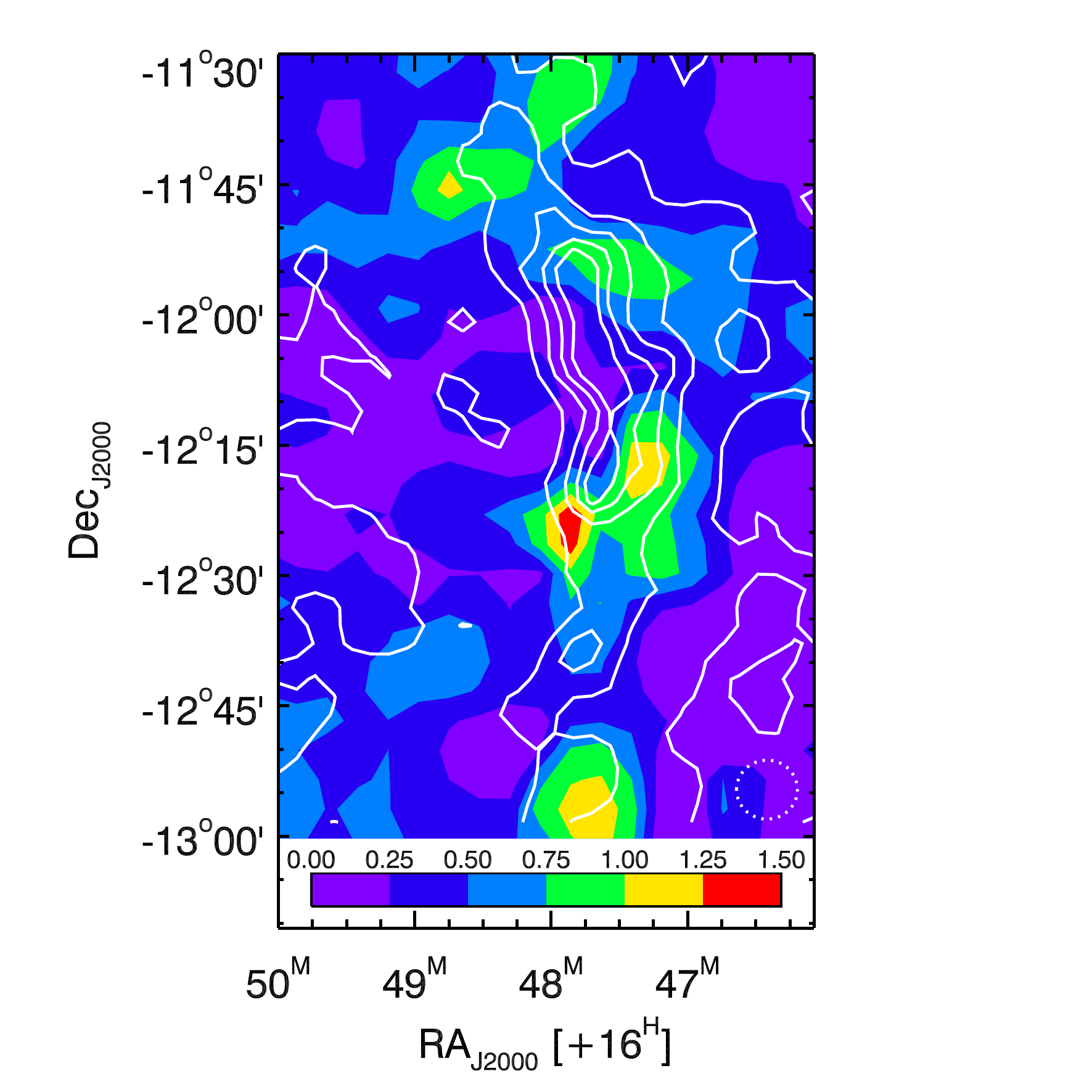}
	\caption{\label{mean_P}Mean, Ricean-corrected percentage polarization $P$ of the Cloud~3 region, computed from mean weighted Stokes $U$ and $Q$ values of BF = 2 (background only) stars. Color levels are stepped linearly from  $0\%$ to $1.5\%$. The NICER extinctions are overlaid as white contours (at the same $A_{\rm V}$ values as used for the gray solid levels in Figure~\ref{extinction}) to show the location of Cloud~3. A dotted white circle in the lower right shows the effective resolution (FWHM) of the $P$ map. (A color version of this figure is available in the online version.)}
\end{figure}

Figure~\ref{mean_P} shows the resulting debiased $P$ map as filled color contours. The white contour lines represent the NICER extinctions and are drawn at the same levels as used for the filled gray levels in Figure~\ref{extinction}. The dotted white circle in the lower right shows the effective resolution (FWHM) of the $P$ map. The highest mean polarization percentages occur on the southern, western, and northern edges of the highest extinction ($A_{\rm V} \geq 6.5$ mag) region. A secondary $P$ enhancement occurs at the southern map boundary. This $P$ map will be discussed in Section 5.1.

\begin{figure}
	\centering
	\includegraphics[width=0.45\textwidth, angle=0, trim = 0.6cm 0cm 3.6cm 0.8cm, clip=true]{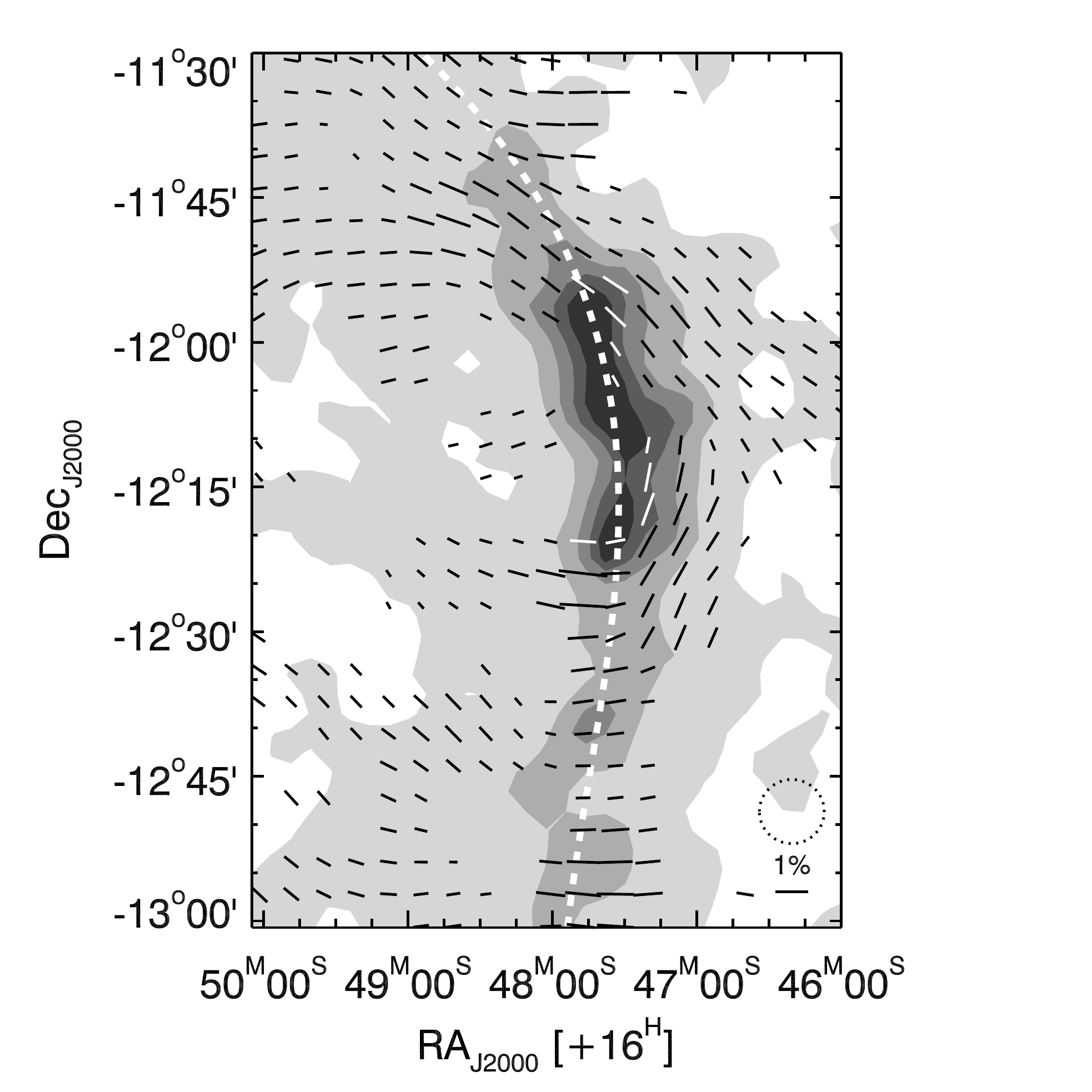}
	\caption{\label{mean_props_all}Grayscale extinction map of Figure~\ref{extinction} with overlaid vectors representing the mean percentage polarization $P$ and position angle P.A. at each grid point, via vector lengths and orientations, respectively. Only grid points with $P/\sigma_{\rm P} \geq 2.5$ are displayed. A dotted black circle in the lower right shows the effective resolution (FWHM) of the polarization map. Below it, a reference 1\% vector is shown. The extinction map uses the same levels as in Figure~\ref{extinction}. The thick, white dashed line represents the polynomial representation of the locations of the extinction ridge of the cloud.}
\end{figure}

Figure~\ref{mean_props_all} shows the NICER extinction grayscale map from Figure~\ref{extinction} overlaid with vectors representing the mean $P$ and P.A. calculated for the high significance bins (those exhibiting $P/\sigma_{\rm P} \geq 2.5$, equivalent to $\sigma_{\rm P.A.} \leq 11.5\degr$). A 1\% vector is shown in the lower right for reference. The thick, white dashed line shows the polynomial representation of the locations of the ridge of the cloud, as described above. The polarization orientations along the southern ridge of the cloud appear to be predominantly perpendicular to the local ridge orientation, while they appear to be predominantly parallel along the northern ridge.

\subsubsection{Ridge Magnetic Field Directions}
\begin{figure}
	\centering
	\includegraphics[width=0.45\textwidth, angle=0, trim = 0cm 1.2cm 0cm 1.4cm, clip=true]{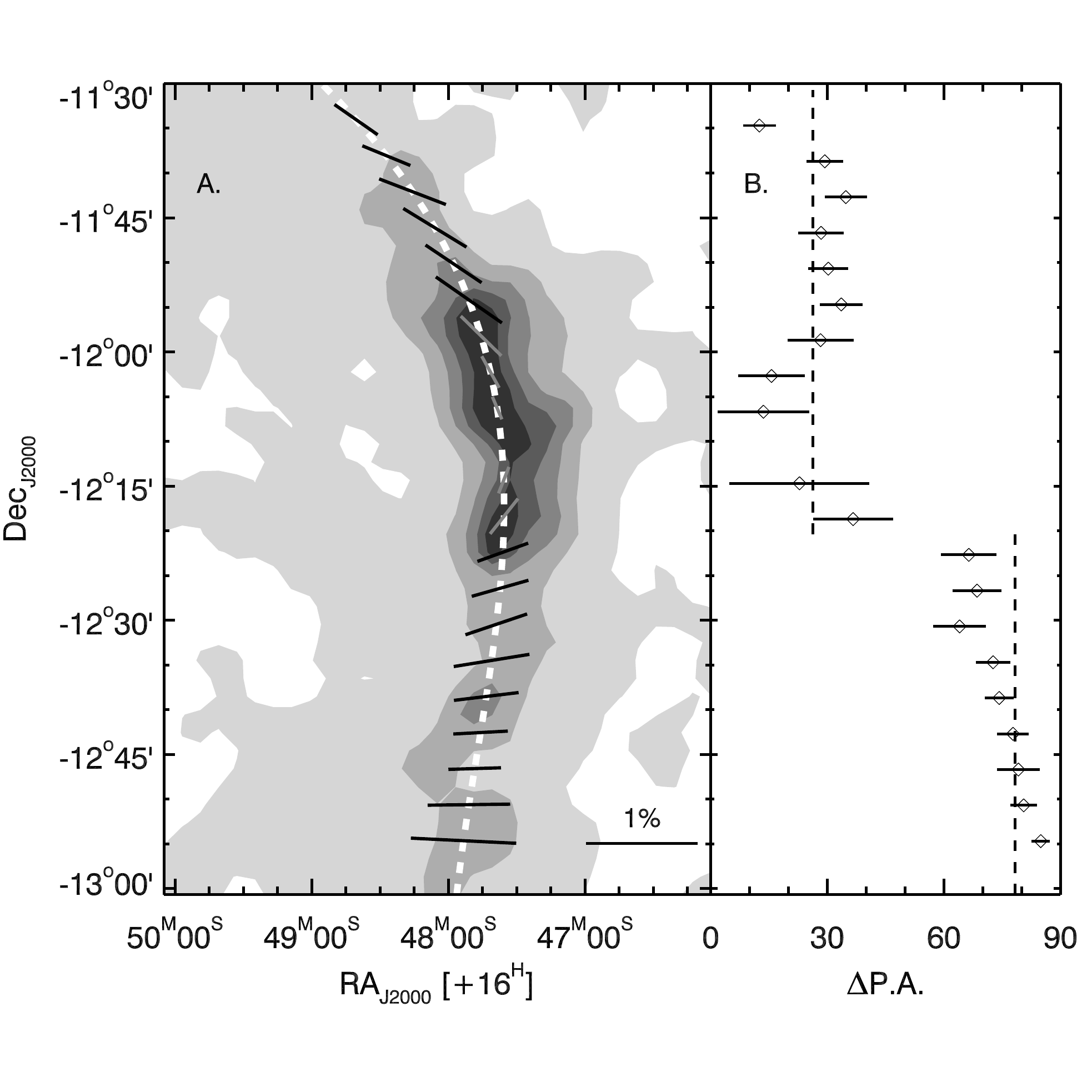}
	\caption{\label{delta_ridge} Polarization properties along the extinction ridge of Cloud~3. (A.) The mean $P$ and P.A. along the ridge are shown as vectors lengths and orientations, respectively. The white dashed line represents the extinction ridge. The extinction map is shown in the background as gray filled regions. (B.) The absolute differences ($\Delta \rm P.A.$) between $\rm P.A._{POL}$ and $\rm P.A._{RIDGE}$ versus decl. The black dashed lines represent the mean $\Delta \rm P.A.s$ of the northern (decl. $\geq -12\degr 20^{\prime}$) and southern (decl. $< -12\degr 20^{\prime}$) populations. Error bars represent the uncertainties in $\rm P.A._{POL}$, only.}
\end{figure}

To further test this observation of field-ridge perpendicularity in the southern region of Cloud~3, the polarization properties along the ridge were examined. Stokes $U$ and $Q$ averaging was performed at RA positions along the ridge using the same bin sizes that were used in the previous analysis and a declination spacing of $4\arcmin$. The ridge locations, resulting polarizations $P$, and polarization P.A.s ($\rm P.A._{POL}$) are shown in the left panel of Figure~\ref{delta_ridge}.

$\rm P.A._{POL}$ was then compared to the position angle of the ridge ($\rm P.A._{RIDGE}$), which was determined at each declination using the tangent to the ridge locations. The differences between the angles ($\Delta \rm P.A. = |\rm P.A._{POL} - \rm P.A._{RIDGE}|$) are shown versus decl. in the right panel of Figure~\ref{delta_ridge}. The error bars for $\Delta \rm P.A.$ represent only the uncertainties in $\rm P.A._{POL}$. Values for $\Delta \rm P.A.$ of $90\degr$ ($0\degr$) indicate that the polarization, and thus the magnetic field, is perpendicular (parallel) to the cloud ridge tangent in those places. The black dashed lines in the right panel of Figure~\ref{delta_ridge} represent the mean $\Delta \rm P.A.s$ of the northern (decl. $\geq -12\degr 20^{\prime}$) and southern (decl. $< -12\degr 20^{\prime}$) populations. In the southern region of Cloud~3, the weighted mean $\Delta \rm P.A.$ is $78 \pm 6\degr$, while the weighted mean $\Delta \rm P.A.$ is $26 \pm 8\degr$ in the northern region of the cloud. This may indicate a transition from a strong-field to a weak-field regime and will be further discussed in Section~5.1.

\subsubsection{Application of Chandrasekhar-Fermi Method}
An estimate of the plane-of-sky (P.O.S.) magnetic field can be obtained from background starlight polarimetry using the Chandrasekhar-Fermi (C-F) method \citep{Chandrasekhar53}. The C-F method, modified by \citet{Ostriker01}, estimates the P.O.S. magnetic field as:
\begin{equation} \label{eq:CF}
B_{\rm P.O.S.} = 1.8 \: \rho^{0.5} \: \frac{\delta v_{\rm L.O.S.}}{\delta \phi}; \delta \phi \leq 25\degr,
\end{equation}
where $B_{\rm P.O.S.}$ is measured in Gauss, $\rho$ is the mean mass density (in $\rm g \, cm^{-3}$), $\delta v_{\rm L.O.S.}$ is the line-of-sight gas velocity dispersion (in $\rm cm \, s^{-1}$), and $\delta \phi$ is the background starlight linear polarization P.A. dispersion (in radians).

To create a full-cloud magnetic field map, $\delta \phi$ was calculated across the cloud using the same bins and weighting as was used for the Stokes $U$ and $Q$ maps. Only BF = 2 stars were used for these calculations. A $\sigma_{\rm P.A.} \leq 15\degr$ cut retained at least 10 stellar polarizations in each bin across the cloud. Due to the $180\degr$ ambiguity of polarization P.A.s (i.e., P.A. = $0\degr$ and $180\degr$ indicate the same magnetic field orientation), dealiasing the P.A.s was sometimes necessary. This dealiasing was done by finding the biased range which produced the minimum $\delta \phi$. P.A. distributions properly centered within biased ranges showed $\delta \phi$ values smaller than when aliasing was present. At each grid point, the minimum $\delta \phi$ was found by sweeping through different biased ranges, in bias increments of $10\degr$. The resulting $\delta \phi$ map had a minimum dispersion of $15.7\degr$ and a median dispersion of $38.5\degr$. Thus, $\delta \phi$ is greater than $25\degr$ for the majority of Cloud~3, making the C-F method inapplicable, as noted by \citet{Ostriker01}.

\begin{figure}
	\centering
	\includegraphics[width=0.45\textwidth, angle=0, trim = 0.6cm 0cm 3.3cm 0.8cm, clip=true]{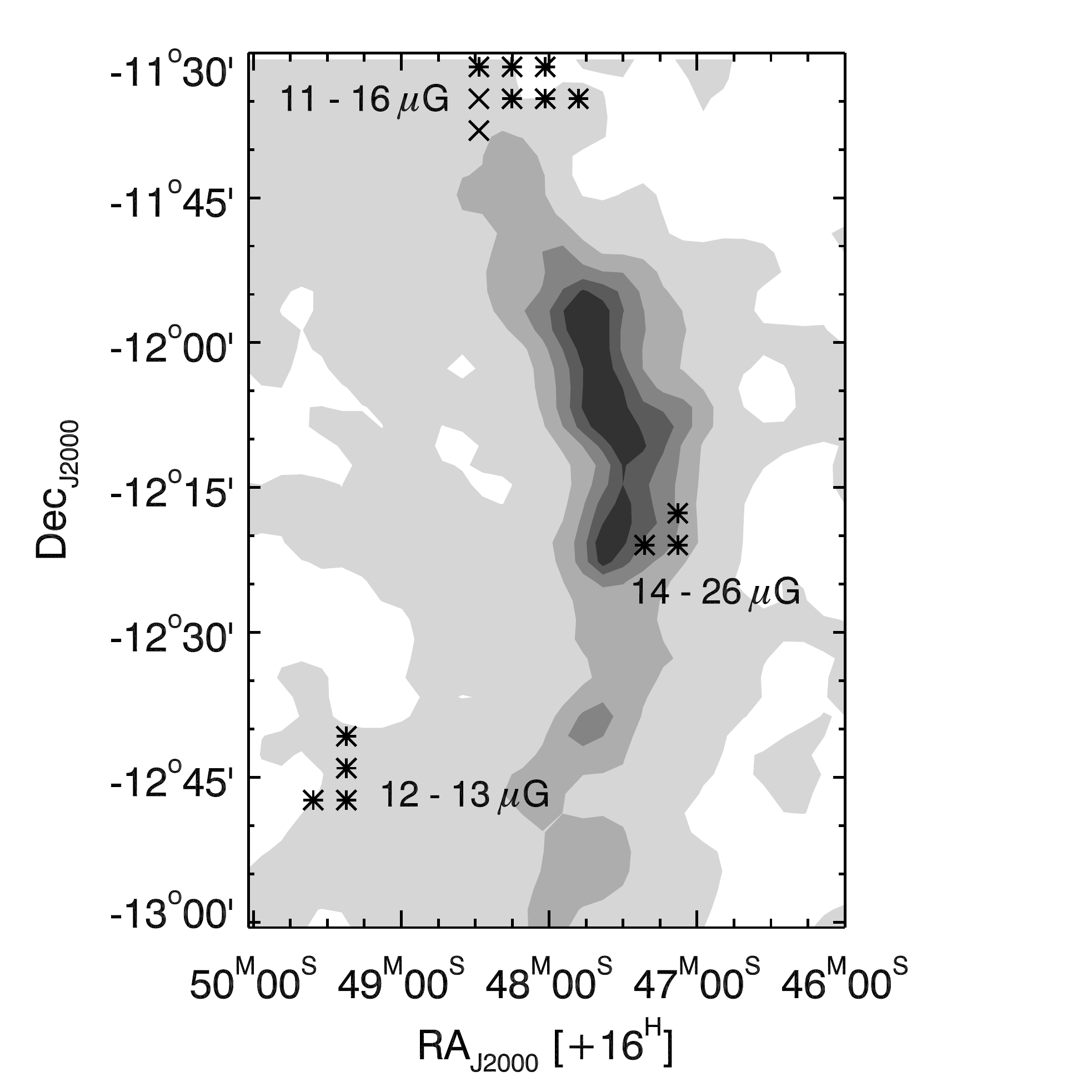}
	\caption{\label{CF}Grayscale extinction map of Figure~\ref{extinction} with average C-F $B_{\rm P.O.S.}$ estimates for three different regions. Each lower (upper) average $B_{\rm P.O.S.}$ estimate was derived using the individual full-width at zero-power (half-power) derived cylindrical model diameters. Black `X's indicate the positions that were used in the full-width at zero-power estimates, while crosses were used in the full-width at half-power ones. The extinction map uses the same levels as in Figure~\ref{extinction}.}
\end{figure}

However, for the 15 bins meeting the \citet{Ostriker01} requirement of $\delta \phi \le 25\degr$ and having $^{13}$CO measurements, an estimate of the plane-of-sky magnetic field could be made. The average $^{13}$CO line profile for each L.O.S. was used to determine $\rho$ and $\delta v_{\rm L.O.S.}$. The H$_2$ column density (in cm$^{-2}$) was first found using:
\begin{equation} \label{eq:NH2}
N_{\rm H_2} = (4.92 \times 10^{20}) \: T \: \sigma_v
\end{equation}
from \citet{Simon01}, where $T$ is the peak antenna temperature in K and $\sigma_v$ is the velocity FWHM. To convert this column density to a volume density, two different effective diameters were estimated, the full-width at half-power (FWHP) and the full-width at zero-power (FWZP). Assuming cylindrical symmetry, $B_{\rm P.O.S.}$ estimates were determined for all 15 bins using the FWZP diameters and 13 bins using the FWHP diameters (the other 2 bins lay outside the FWHP radius).

The 15 bins are located in three distinct regions, so an average estimate of $B_{\rm P.O.S.}$ (in~$\mu$G) was calculated for each region, as shown in Figure~\ref{CF}. The lower average estimates ($\sim 11 - 14$~$\mu$G) are derived from the FWZP diameters, which are larger and thus yield lower values of $\rho$. Within each region, the standard deviation of $B_{\rm P.O.S.}$ is $\sim 2$~$\mu$G. The bins used in these calculations are denoted by black `X's. The higher average estimates ($\sim 13 - 26$~$\mu$G) are derived from the FWHP diameters.  The standard deviation within each of the three regions is $\sim 3$~$\mu$G. The bins used in these calculations are denoted by the black crosses, which lie on top of the `X's.

These average estimates of $B_{\rm P.O.S.}$ reveal a moderate strength magnetic field, which is consistent with the total field strength estimate of \citet{Heiles88}, but are much smaller than the \citet{McCutcheon86} estimate. These new $B_{\rm P.O.S.}$ estimates also indicate that the field strength may be somewhat greater in the dense core of the cloud.

\begin{figure}
	\centering
	\includegraphics[width=0.45\textwidth, angle=0, trim = 0cm 0cm 0.7cm 0.1cm, clip=true]{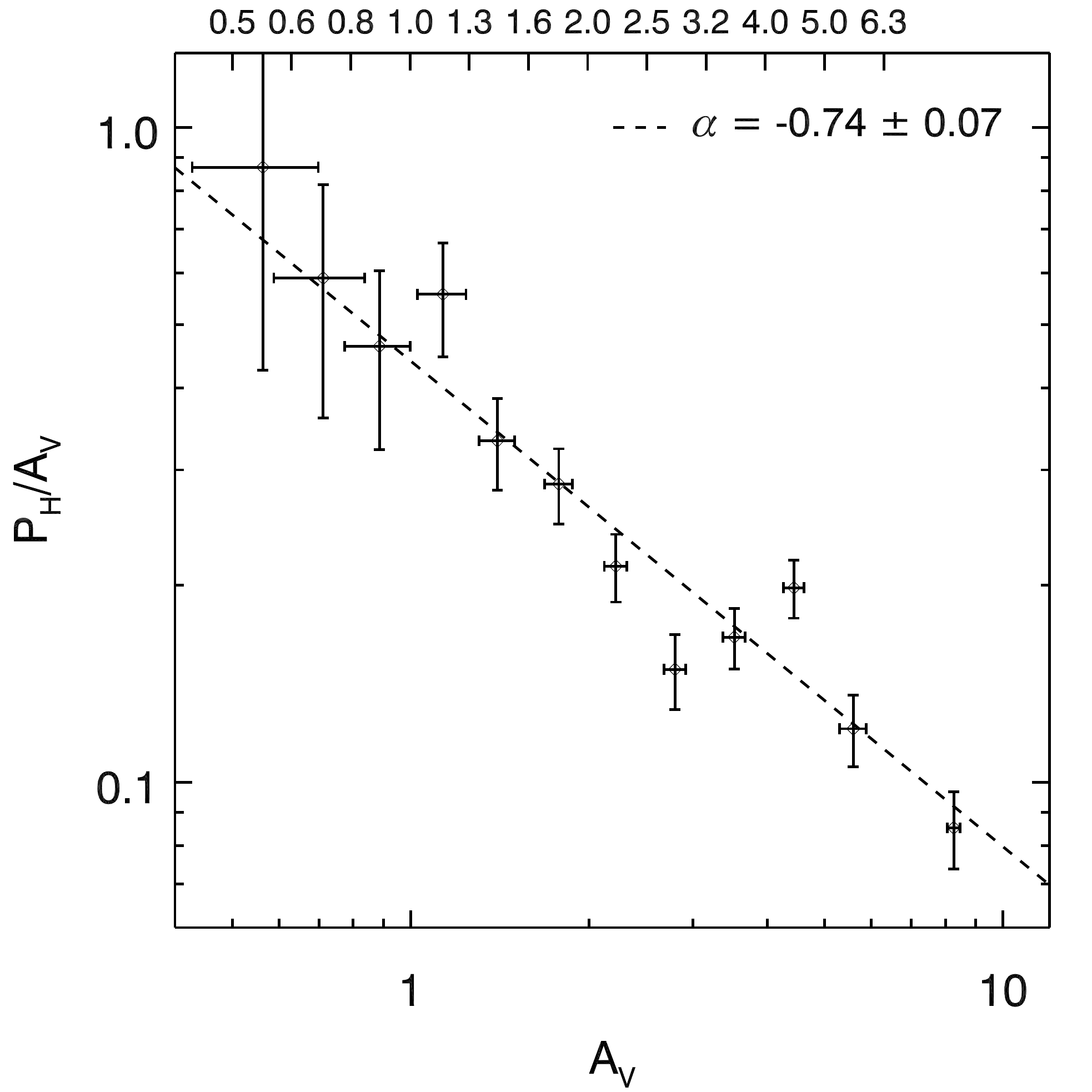}
	\caption{\label{PE}Mean polarization efficiency ($P_{\rm H}/A_{\rm V}$) versus logarithmically-binned mean NICER-traced extinctions ($A_{\rm V}$) for the Cloud~3 region. The top axis shows the boundaries of the extinction bins. Error bars represent the $1\sigma$ uncertainties in the means. Some low $A_{\rm V}$ bins have mean $A_{\rm V}$ uncertainties wider than their bin widths. The dashed line is a least-squares power-law fit to the data.}
\end{figure}

\subsection{Polarization Efficiency}

The mean $A_{\rm V}$ through the Cloud~3 region is low ($\sim$ 0.5 mag) and should allow for RAT aligning radiation to penetrate into the cloud. To study grain alignment efficiency in L204, the polarization efficiency\footnote{Note that this definition is a mixed wavelength estimator of PE and was chosen in an effort to stay close to the observed properties, with the least invocation of dust or polarization models.} (PE $\equiv P_{\rm H}/A_{\rm V}$) was examined, as a function of $A_{\rm V}$, for the entire surveyed region. To increase signal to noise, background stars (BF = 2) with UF = 1 and 2 were placed into 12 logarithmically-spaced extinction bins, based on their nominal $A_{\rm V}$ values (from Table~\ref{stars}). Weighted mean PE and $A_{\rm V}$ values were calculated using the nominal values, variance-weighted by their uncertainties. Bin uncertainties for $A_{\rm V}$ and PE were computed from the summed variance weights. For some low $A_{\rm V}$ bins, these bin uncertainties exceed the bin widths. Figure~\ref{PE} shows the resulting $\langle A_{\rm V} \rangle$ and $\langle \rm PE \rangle$, as well as the bin boundaries (top axis). A power-law index was found by linear regression to:
\begin{equation} \label{eq:one}
log(\frac{P_{\rm H}}{A_{\rm V}}) = \alpha \: log(A_{\rm V})+\beta,
\end{equation}
where $\alpha$ is the power-law index and $\beta$ is a constant offset. The dashed line in Figure~\ref{PE} represents the least-squares fit, with an $\alpha$ index of $-0.74 \pm 0.07$.

\begin{figure}
	\centering
	\includegraphics[width=0.45\textwidth, angle=0, trim = 0.5cm 0.2cm 3.2cm 0.9cm, clip=true]{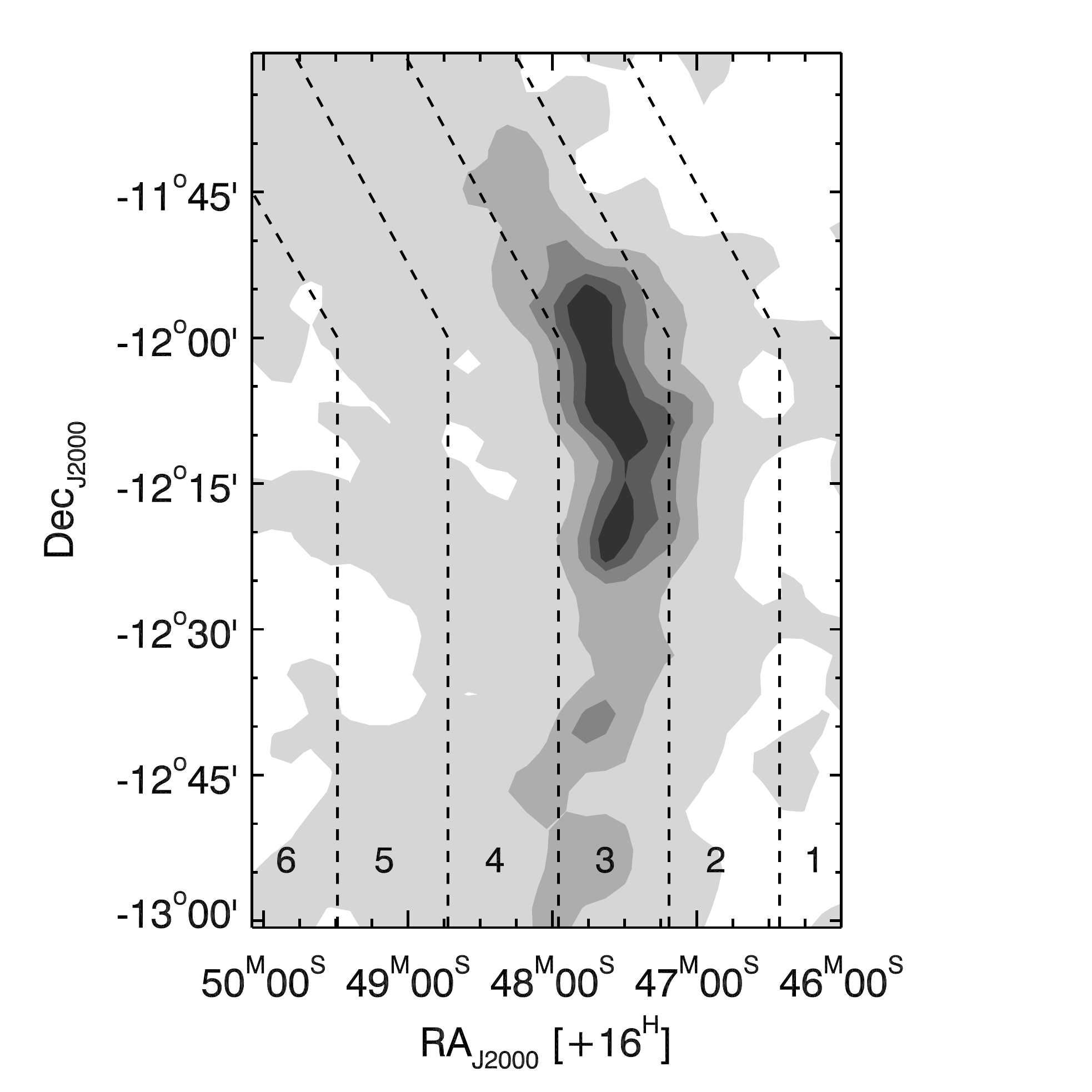}
	\caption{\label{regions} Regions for mean polarization efficiency PE determination. Region boundaries were chosen to reflect the shape of Cloud~3 and to test for changes in polarization efficiency with distance from $\zeta$~Oph. Region numbers are indicated at the bottom. Figure~\ref{extinction} extinctions are shown in gray-scale.}
\end{figure}

However, there may be problems with this technique. As pointed out by \citet{Andersson07}, the extinctions used are representative of the optical depth ${\it along}$ the line of sight to each background star, but more important is the optical depth ${\it between}$ the illuminator and the aligned dust grains, which is harder to measure. To try to account for this ``true'' optical depth, PE was also examined as a function of distance from $\zeta$~Oph. The surveyed region was divided into six regions, shown in Figure~\ref{regions}. The region boundaries were chosen based on RA (i.e., distance from $\zeta$~Oph) and the shape of cloud, determined by examining the extinction map. The regions are labeled ``1''-``6'', with larger region numbers corresponding to larger angular offsets from $\zeta$~Oph.

\begin{figure}
	\centering
	\includegraphics[width=0.45\textwidth, angle=0, trim = 1.2cm 0.1cm 0.9cm .1cm, clip=true]{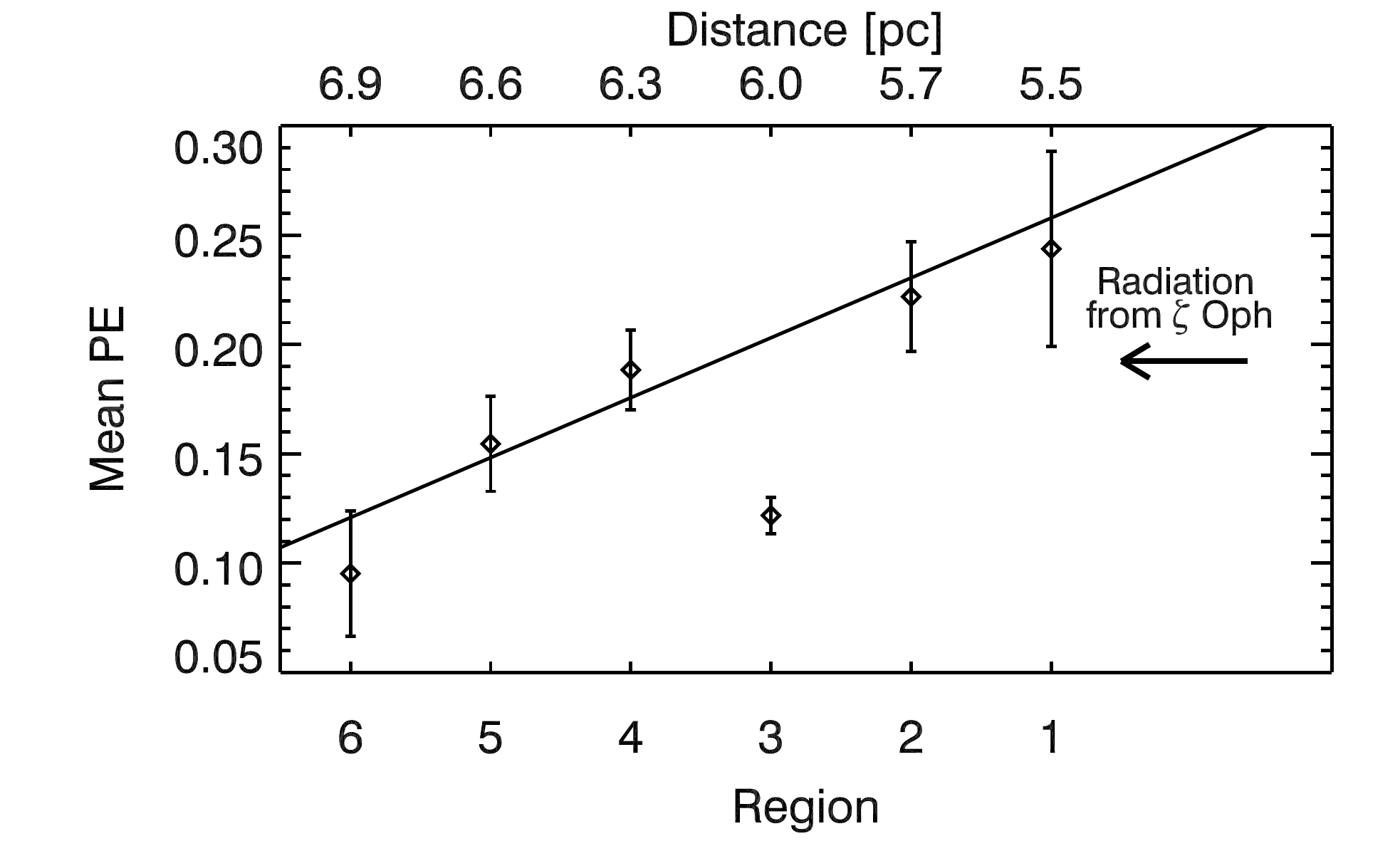}
	\caption{\label{pol_eff} Weighted mean polarization efficiencies (${\rm PE} \equiv P_{\rm H}/A_{\rm V}$), calculated for each region defined in Figure~\ref{regions}. Error bars show $1\sigma$ uncertainties in the means. The black solid line is a linear fit to the run of mean PE, excluding Region~3, and shows a slope of $-0.027 \pm 0.008$~\% polarization per mag per region. The top axis gives the distance from $\zeta$~Oph, assuming Cloud~3 and $\zeta$~Oph are at the same 112~pc distance.}
\end{figure}

The weighted mean PE and its uncertainty were calculated within each of these regions, and the resulting values are shown in Figure~\ref{pol_eff}. With the exception of Region~3, the center of the cloud, the mean PE is well fit by a linear regression with a slope of $-0.027 \pm 0.008$~\% polarization per mag per region (or $-0.09 \pm 0.03 \% \, {\rm mag}^{-1} \, {\rm pc}^{-1}$). This indicates that as the distance from $\zeta$~Oph increases, the polarization efficiency decreases. Additionally, in the core of the cloud, which is well shielded from radiation, the mean polarization efficiency is $9.8\sigma$ less than that predicted by the fit to the rest of the cloud. (This difference decreases to $2\sigma$ when also taking into account the uncertainties in the fitted line coefficients.)

\begin{figure}
	\centering
	\includegraphics[width=0.45\textwidth, angle=0, trim = 1.2cm 0.1cm 0.9cm .1cm, clip=true]{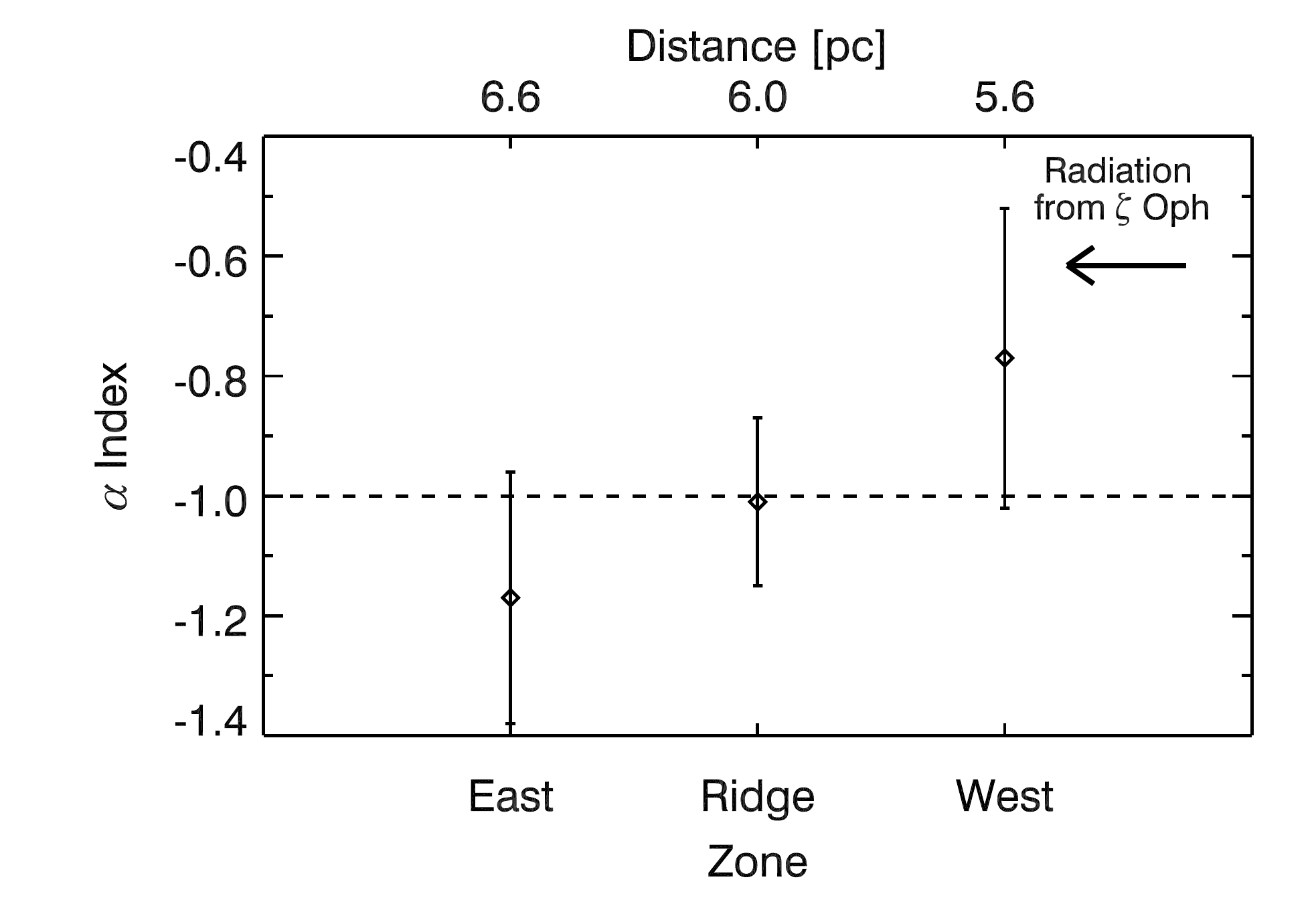}
	\caption{\label{PE_fits} The PE power-law $\alpha$ index versus zone in the survey region. Error bars represent $1\sigma$ uncertainties in the fitted $\alpha$ indices. An $\alpha$ index of $-1$ (black dashed line) is expected if the distributions of $P_{\rm H}$ and $A_{\rm V}$ are fully uncorrelated.}
\end{figure}

To further test for shadowing effects in Cloud~3, stars in the field were assigned to one of three larger zones: the ridge (Region~3), the west side of the cloud (Regions 1 \& 2), and the east side (Regions 4-6). Within each zone, the stars were placed into the same 12 logarithmically-spaced extinction bins used in Figure~\ref{PE}. Linear, least-squares fits of $\langle PE \rangle $ vs $\langle A_{\rm V} \rangle$ were calculated in each of the three zones, using only bins with at least 5 stars and Equation~\ref{eq:one}, to find the power-law index $\alpha$ in each zone. The $\alpha$ indices and their uncertainties are shown in Figure~\ref{PE_fits}. A power-law index of $\alpha = -1$, shown by the black dashed line, is characteristic of uncorrelated polarizations and extinctions (i.e., forced to be $-$1 because PE $\propto 1/A_{\rm V}$).

The $\alpha$ index changes across the cloud and appears to decrease with distance from $\zeta$~Oph. The $\alpha$ index in the western zone closely matches that seen for the entire Cloud~3 region in Figure~\ref{PE} ($-0.77 \pm 0.25$ vs $-0.74 \pm 0.07$). This suggests that the stars in the western zone may be dominating the power-law fit for the entire cloud. Due to the small number of stars in each zone, the uncertainties in their $\alpha$ indices are significant and none of the zones differ significantly from $\alpha = -1$. Taken at face value, even though the mean PE decreases with distance from $\zeta$~Oph, the PE slope with $A_{\rm V}$ has not been found to vary significantly with that distance. These results will be further discussed in Section 5.2.

\section{Discussion}
The NIR background starlight polarizations observed towards Cloud~3 of L204 have revealed a wealth of information about conditions within and around the cloud. Through Stokes $U$ and $Q$ averaging of the many background starlight polarizations, including those with low individual significance, the magnetic field structure of Cloud~3 has been revealed. Combined with NICER extinctions, the stellar polarizations also provided insight into the dust grain alignment efficiency throughout the cloud. In the following sections, these properties are highlighted.

\subsection{Magnetic Field Characteristics}
Previous optical polarimetric observations of stars behind L204 \citep{McCutcheon86} revealed that the magnetic field is predominantly perpendicular to the long axis of the cloud complex and is coherent over separations of several parsecs. The new observations presented here, however, reveal a more complex magnetic field on smaller scales. The mean polarization percentage map in Figure~\ref{mean_P} does reveal that larger polarizations tend to be associated with regions of modest extinction, implying that the magnetic field being probed is associated with the cloud and is aligning internal dust grains.

\begin{figure*}
	\centering
	\includegraphics[width=0.8\textwidth, angle=0, trim = 0.5cm 3.9cm 0.9cm 0.9cm, clip=true]{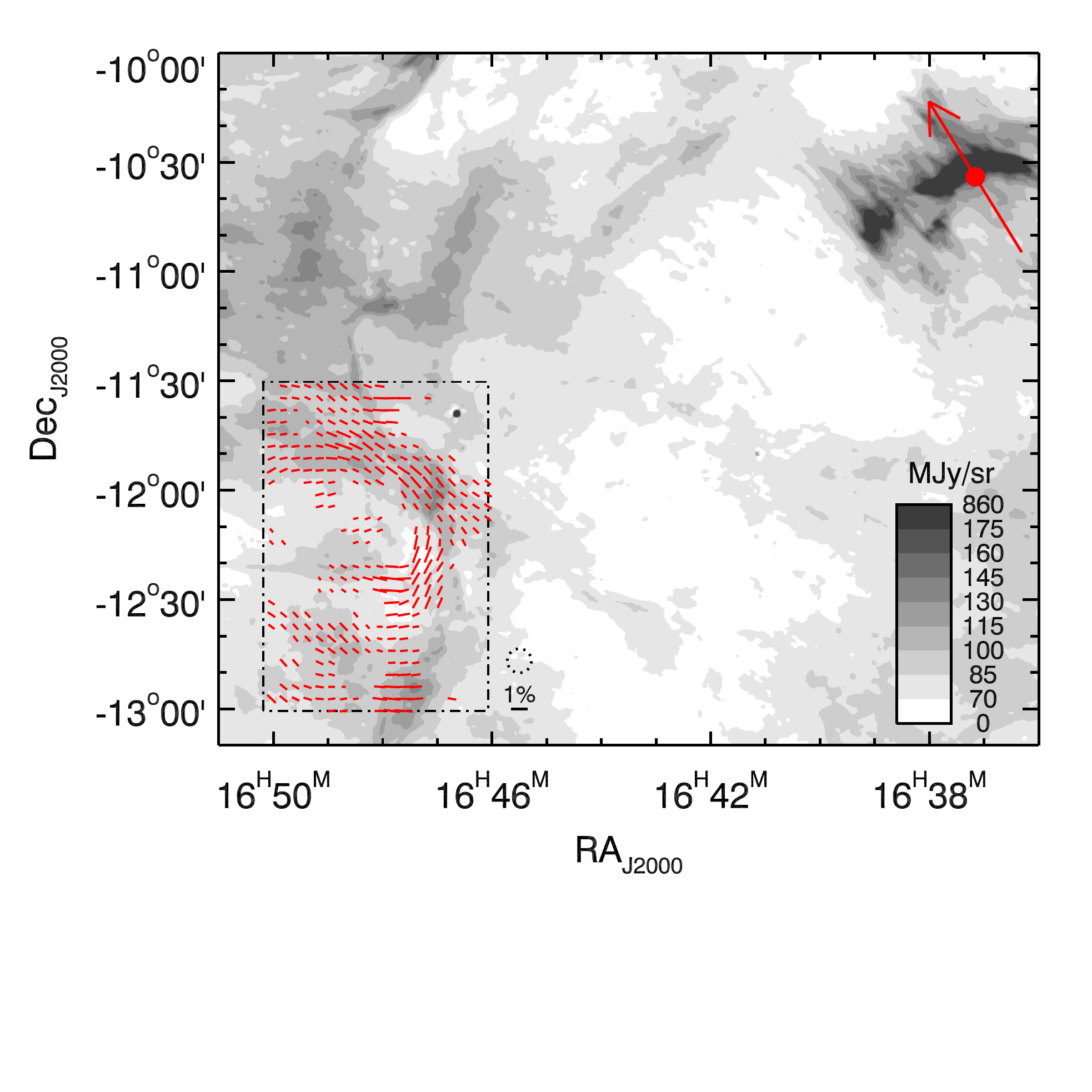}
	\caption{\label{pol_lab} The L204 Region from Figure~\ref{lab}. The polarization map from Figure~\ref{mean_props_all} is overlaid as red vectors. A dotted black circle to the lower right of the observed region (black dot-dashed line) shows the effective resolution (FWHM) of the polarization map. Below it, a reference 1\% vector is shown. The magnetic field and dust in the northern and western regions appear to be compressed against the $\zeta$~Oph facing side of Cloud~3. (A color version of this figure is available in the online version.)}
\end{figure*}

Figure~\ref{pol_lab} shows a similar view of the L204 region as that in Figure~\ref{lab}, overlaid in red by the mean polarization map from Figure~\ref{mean_props_all}. This map reveals a wide range of P.A.s in the vicinity of Cloud~3. Though there are changes in P.A. within and across the cloud, the direction of the magnetic field is predominantly perpendicular to the long axis of the cloud in the southern region, as seen in Figure~\ref{delta_ridge}. This axis-perpendicular field, which is shielded by the rest of the cloud and less likely to have been influenced by $\zeta$~Oph, may be a remnant signature of the collapse of L204. This would indicate that L204 collapsed under strong-field conditions, e.g. along the magnetic field lines.

However, at the northern and western edge of the cloud (facing $\zeta$~Oph), the P.A.s are parallel to the extended 12$\mu$m dust emission, appearing to be compressed against the side of the cloud. This may indicate that while clearing out a cavity, the UV flux from $\zeta$~Oph has pushed the magnetic field on the western side of L204 up against the high density core. This discrepancy between the northern and southern regions of Cloud~3 may be a sign of a transition between strong-field collapse and present weak-field conditions.

Though the P.A.s appear to vary smoothly in Figure~\ref{pol_lab}, the dispersions of local sets of individual stellar polarizations are large, up to $50\degr$ in some parts of the cloud. As discussed in Section 4.2, these large dispersions are unsuitable for estimating the P.O.S. magnetic field strengths for most of the cloud using the C-F method (which seems more applicable to quiescent conditions, e.g., \citealt{Marchwinski12}). However, an estimate of the P.O.S. magnetic field strength was possible for 15 positions and yielded an average $B_{\rm P.O.S.} \sim 11 - 26$~$\mu$G. These estimates are consistent with the estimates of \citet{Heiles88}, but smaller than the \citep{McCutcheon86} estimated value. The large P.A. dispersions found in Cloud~3 suggest that there may be significant turbulent motions and tangled magnetic fields present. \citet{Tachihara12} performed CO $J = 1-0$ and $3-2$ observations of the western edge of Cloud~3 and found evidence of turbulence driven by thermal instability.

\subsection{Grain Alignment}
Polarization efficiency, and grain alignment efficiency, are known to decrease in the dense cores of molecular clouds \citep{Goodman95}. This decrease is predicted by many grain alignment theories, including RAT theory. A unique prediction of RAT theory \citep{Lazarian07}, however, is that in the presence of a strong, anisotropic radiation field, dust grains will remain aligned deep into molecular clouds.

For the entire surveyed Cloud~3 region, the PE versus $A_{\rm V}$ fit power-law index $\alpha$, shown in Figure~\ref{PE}, is $-0.74 \pm 0.07$. This is significantly steeper than similar indices found for other clouds. In the Taurus Molecular Cloud, \citet{Gerakines95} found an index of $-0.56 \pm 0.17$ at $K$-band. \citet{Whittet08} found indices of $-0.52 \pm 0.07$ and $-0.45 \pm 0.10$ for the Taurus and Ophiucus clouds, respectively, at $K$-band. Together, these imply that grain alignment in Cloud~3 in L204 becomes ${\it less}$ efficient as extinction increases than in the other clouds. Thus, it would appear that the strong radiation field from $\zeta$~Oph is less able to boost grain alignment efficiency in L204, relative to these other clouds.

However, as mentioned in Section 4.3, the extinctions used are representative of the optical depth ${\it along}$ the L.O.S. to background stars and are not providing the optical depth between $\zeta$~Oph and the dust. Figure~\ref{pol_eff} tries to take this effect into account by testing how the PE changes with distance from $\zeta$~Oph. With the exception of Region~3, the PE steadily ${\it decreases}$ with increasing distance from $\zeta$~Oph. This relation indicates that where they are exposed to a stronger radiation field, dust grains are more efficiently aligned.

The unusually low mean polarization efficiency in Region~3 is not unexpected, because this region contains the highly extincted Cloud~3 core. As seen in Figure~\ref{WISE}, the core of the cloud (RA $\simeq$ 16:47:30, decl. $\simeq -$12:15:00) is opaque even at 4.6$\mu$m, making it unlikely that short $\lambda$ radiation, from $\zeta$~Oph or elsewhere, is reaching these grains, and so prevents background starlight from finding aligned grains to easily reveal any magnetic field there.

Further evidence of a possible change in grain alignment efficiency with distance from $\zeta$~Oph was shown in Figure~\ref{PE_fits}. Though all three power-law indices are marginally consistent with the polarization being independent of extinction, there remains an apparent change in $\alpha$ with distance from $\zeta$~Oph. On the $\zeta$~Oph facing (West) side of Cloud~3, the $\alpha$ index is flatter, suggesting dust grains remain better aligned to higher column densities. Farther away from $\zeta$~Oph (East), the $\alpha$ index is steeper, indicating less polarization per mag of $A_{\rm V}$. A linear fit to the data in Figure~\ref{PE_fits} (not shown) produces a slope of $0.20 \pm 0.16$. This, however, is not a significant change and does not exclude a constant $\alpha$ index throughout the cloud.

As with Figure~\ref{PE}, the polarization efficiencies in Figure~\ref{pol_eff} and Figure~\ref{PE_fits} were calculated using L.O.S. extinctions. Though these are not strictly correct optical depths, they are readily available. Truer estimates of the optical depths between $\zeta$~Oph and the aligned dust grains may reveal stronger correlations.

Due to the projection of L204 in front the HII region surrounding $\zeta$~Oph, the relative placements of L204 and $\zeta$~Oph are known, but the exact geometry of the system is unknown. The decay in PE with offset from $\zeta$~Oph in Figure~\ref{pol_eff} falls far faster than r$^{-2}$, even assuming $\zeta$~Oph and Cloud~3 are at the same distance. Thus, distance dilution of the radiation field alone is insufficient to account for the PE decay and another factor, such as extinction, has a lead role. This implies that the assumption that L204 and $\zeta$~Oph are at the same distance is not a bad one and that the eastern side of the cloud is being shielded by the western side.

\subsubsection{V446~Oph}
Due to its bright appearance in Figure~\ref{WISE}, one might conclude that V446~Oph, not $\zeta$~Oph, is illuminating Cloud~3. However, V446~Oph is probably much more distant and unlikely to interact with L204. It is a semiregular variable star with a $B$-band amplitude of $\sim 1$ mag \citep{Samus09}, has a spectral type of M8 \citep{Hansen75}, and hosts an SiO maser \citep{Hall90}. Based on these, V446~Oph is likely a red supergiant. If an absolute $K_{\rm S}$ magnitude of $-$6.5 mag is assumed for all M I-II stars \citep{Nikolaev00}, then with $K_{\rm S} = 0.923$ mag \citep{Skrutskie06}, V446~Oph would have to experience 24 mag of visual extinction to be at the same distance as L204. This is inconsistent with its NICER estimated extinction of $A_{\rm V} = 5.5 \pm 2.6$ mag, which places it at a minimum distance of $\sim 243$~pc. This is well beyond the L204 complex, reaffirming that V446~Oph is not the illuminator for L204.

\section{Summary}

We have used NIR $H$-band linear polarization of background starlight to probe the magnetic field of Cloud~3 in L204. The mean polarization properties of the cloud were determined through spatial averaging of the Stokes $U$ and $Q$ values for the 3,896 background stars with measured polarizations and exhibiting $A_{\rm V} \ge 0.5$ mag.

These mean properties revealed that Cloud~3 has a coherent magnetic field, oriented predominantly perpendicular to the long axis of the cloud in its southern region. There, the mean deviation of the field from being cloud-perpendicular is $12\degr$. The small deviations from cloud-perpendicular in the southern portion of the main cloud ridge support strong magnetic field models and suggest that this portion of the cloud collapsed (or has been pushed or driven) along the magnetic field lines. The corresponding mean deviation of the field from the cloud long axis in the northern region is $64\degr$. The direction of the field in this region implies that the UV flux from $\zeta$~Oph has compressed the (weak) magnetic field against the face of Cloud~3.

The $B_{\rm P.O.S.}$ field strength is estimated to be $\sim 11 - 26$ $\mu$G, using the C-F technique, for the small number of locations exhibiting sufficiently small P.A. dispersions.

The dust-grain alignment was probed via measuring the $P_{\rm H}/A_{\rm V}$ vs $A_{\rm V}$ power-law index $\alpha$. For the entire Cloud~3 region, this index is $-0.74 \pm 0.07$, which is steeper (i.e., shows less grain alignment at high extinctions) than similar indices found by others for other molecular clouds. However, the polarization efficiency and, thus, dust grain alignment efficiency changes with location and extinction within Cloud~3. Specifically, grain alignment efficiency decreases with increasing distance from $\zeta$~Oph and is also significantly lower in the dense core of Cloud~3.

\acknowledgements
The authors thank M. Pavel and A. Pinnick for performing some of the observations and for helpful discussions. The authors thank K. Tachihara for making the mm CO data available. This publication makes use of data products from the Two Micron All Sky Survey, which is a joint project of the University of Massachusetts and the Infrared Processing and Analysis Center (IPAC)/California Institute of Technology (CalTech), funded by NASA and NSF and the from ${\it Wide}$-${\it field}$ ${\it Infrared}$ ${\it Survey}$ ${\it Explorer}$, which is a joint project of the University of California, Los Angeles, and JPL/CalTech, funded by NASA. This research made use of the C. Beaumont IDL library, available at www.ifa.hawaii.edu/users/beaumont/code. This research was conducted in part using the Mimir instrument, jointly developed at Boston University and Lowell Observatory and supported by NASA, NSF, and the W.M. Keck Foundation. This work and the analysis software for Mimir data were developed under NSF grants AST 06-07500 and 09-07790 to Boston University.

{\it Facility:} \facility{Perkins (Mimir)}

\clearpage

\begin{turnpage}
\begin{deluxetable*}{lllllllllllllll}
\tabletypesize{\footnotesize}
\tablecaption{Polarimetry and Photometry of 7150 Stars Towards L204 - Cloud~3\label{stars}}
\tablewidth{0pt} 
\tablehead{
\colhead{} & \multicolumn{8}{c}{Mimir} & \multicolumn{3}{c}{2MASS\tablenotemark{a}} & \colhead{} & \colhead{NICER} & \colhead{}\\
\cline{2-8} \cline{10-12}
\colhead{Star} & \colhead{R.A. [J2000]} & \colhead{Decl. [J2000]} & \colhead{$U$} & \colhead{$Q$} & \colhead{$P$\tablenotemark{b}} & \colhead{P.A.\tablenotemark{c}} & \colhead{UF\tablenotemark{d}} & \colhead{} & \colhead{$J$} & \colhead{$H$} & \colhead{$K$} & \colhead{} & \colhead{$A_{\rm V}$\tablenotemark{e}} & \colhead{BF\tablenotemark{f}}\\
\colhead{} & \colhead{HH MM SS.S} & \colhead{DD MM SS} & \colhead{\%} & \colhead{\%} & \colhead{\%} & \colhead{$\deg$} & \colhead{} & \colhead{} & \colhead{mag} & \colhead{mag} & \colhead{mag} & \colhead{} & \colhead{mag} & \colhead{}\\
\colhead{(1)} & \colhead{(2)} & \colhead{(3)} & \colhead{(4)} & \colhead{(5)} & \colhead{(6)} & \colhead{(7)} & \colhead{(8)} & \colhead{} & \colhead{(9)} & \colhead{(10)} & \colhead{(11)} & \colhead{} & \colhead{(12)} & \colhead{(13)}
}
\startdata
   1 & 16 46 04.3 & -11 34 18 &   8.9 $\pm$  7.8 &  -5.7 $\pm$   7.6 &   7.2 $\pm$   7.7 &   61 $\pm$   30 &   2 &   &  14.23 $\pm$ 0.03 &  13.77 $\pm$ 0.03 &  13.62 $\pm$ 0.04 &   & -0.71 $\pm$ 1.17 &   1\\
   2 & 16 46 04.4 & -11 37 46 &  10 $\pm$  12 &  13 $\pm$  11 &  11 $\pm$  12 &   18 $\pm$   31 &   3 &   &  15.15 $\pm$ 0.04 &  14.31 $\pm$ 0.04 &  14.08 $\pm$ 0.06 &   & 2.00 $\pm$ 1.21 &   2\\
   3 & 16 46 04.8 & -12 04 38 &   1 $\pm$  20 &  -13 $\pm$  20 &   0 $\pm$  20 &    0 $\pm$  180 &   3 &   &  15.13 $\pm$ 0.06 &  14.45 $\pm$ 0.05 &  14.43 $\pm$ 0.09 &   & -0.03 $\pm$ 1.29 &   1\\
   4 & 16 46 04.8 & -11 46 04 &   3.8 $\pm$  7.8 &  -0.0 $\pm$   7.6 &   0.0 $\pm$   7.8 &    0 $\pm$  180 &   3 &   &  14.82 $\pm$ 0.03 &  14.20 $\pm$ 0.03 &  14.08 $\pm$ 0.06 &   & 0.04 $\pm$ 1.20 &   1\\
   5 & 16 46 05.0 & -12 38 31 &   5 $\pm$  13 &   -1 $\pm$  13 &   0 $\pm$  13 &    0 $\pm$  180 &   3 &   &  15.82 $\pm$ 0.09 &  15.31 $\pm$ 0.10 &  15.21 $\pm$ 0.14 &   & -0.65 $\pm$ 1.47 &   1\\
   6 & 16 46 05.1 & -11 45 54 &   2 $\pm$  15 &   -8 $\pm$  14 &   0 $\pm$  14 &    0 $\pm$  180 &   3 &   &  15.73 $\pm$ 0.07 &  14.82 $\pm$ 0.07 &  14.69 $\pm$ 0.10 &   & 2.06 $\pm$ 1.33 &   2\\
   7 & 16 46 05.2 & -12 35 24 &  -31 $\pm$  11 &   -2 $\pm$  12 &  28 $\pm$  11 &  133 $\pm$   12 &   3 &   &  16.27 $\pm$ 0.12 &  15.16 $\pm$ 0.08 &  14.92 $\pm$ 0.11 &   & 3.70 $\pm$ 1.47 &   2\\
   8 & 16 46 05.3 & -12 14 44 &  -1.7 $\pm$  4.8 &   1.2 $\pm$   5.0 &   0.0 $\pm$   4.9 &    0 $\pm$  180 &   2 &   &  13.86 $\pm$ 0.03 &  13.19 $\pm$ 0.03 &  12.95 $\pm$ 0.03 &   & 1.07 $\pm$ 1.16 &   2\\
   9 & 16 46 05.4 & -12 31 38 &   6.6 $\pm$  4.0 &  -2.1 $\pm$   4.2 &   5.6 $\pm$   4.1 &   53 $\pm$   20 &   3 &   &  14.77 $\pm$ 0.04 &  14.16 $\pm$ 0.04 &  13.95 $\pm$ 0.05 &   & 0.56 $\pm$ 1.20 &   2\\
  10 & 16 46 05.5 & -12 16 59 &   -3 $\pm$  10 &  10 $\pm$  11 &   0 $\pm$  11 &    0 $\pm$  180 &   3 &   &  14.88 $\pm$ 0.05 &  14.22 $\pm$ 0.05 &  14.03 $\pm$ 0.07 &   & 0.70 $\pm$ 1.23 &   2\\
\enddata
\tablenotetext{a}{$J$, $H$, and $K$ values of 99.99 indicate that no matching 2MASS star was found. Magnitude uncertainties of 9.99 indicate measurements which are upper limits.}
\tablenotetext{b}{$P$ values have been corrected for positive bias. The correction was also propagated into the P.A. uncertainties.}
\tablenotetext{c}{P.A.s are equatorial, measured East of North.}
\tablenotetext{d}{Usage Flag: 1 = highest quality; 2 = moderate;, 3 = poor}
\tablenotetext{e}{$A_{\rm V}$ values for stars with 2MASS upper limits or without a 2MASS match are set to $-9.99 \pm 0.00$.}
\tablenotetext{f}{Background Flag: 2 = background; 1 = foreground; 0 = unknown}
\end{deluxetable*}
\end{turnpage}

\end{document}